\documentclass[a4paper,11pt]{article}

\usepackage{jheppub}                
\usepackage[all]{hypcap}            
\usepackage{slashed}                
\usepackage[usenames,table]{xcolor} 
\usepackage{feyndiag}               
\usepackage{parskip}                
\usepackage[section]{placeins}      
\usepackage[utf8]{inputenc}         
\usepackage{bbm}                    
\usepackage{ytableau}               
\usepackage{afterpage}              

\renewcommand{\arraystretch}{1.4}
\newcommand{\irrep}[2][0]{\ensuremath{\mathrm{\textbf{#2}}}}

\graphicspath{{./plots/}}
\title{A Sommerfeld Toolbox for Colored Dark Sectors}
\author[a]{Sonia~El~Hedri,}
\author[a]{Anna~Kaminska,}
\author[a]{Maikel~de~Vries}
\affiliation[a]{PRISMA Cluster of Excellence \& Mainz Institute for Theoretical Physics, Johannes Gutenberg University, 55099 Mainz, Germany}
\emailAdd{elhed001@uni-mainz.de}
\emailAdd{akaminsk@uni-mainz.de}
\emailAdd{mdevrie@uni-mainz.de}
\preprint{MITP/16-135}
\arxivnumber{1612.02825}

\abstract{We present analytical formulae for the Sommerfeld corrections to the annihilation of massive colored particles into quarks and gluons through the strong interaction. These corrections are essential to accurately compute the dark matter relic density for coannihilation with colored partners. Our formulae allow to compute the Sommerfeld effect, not only for the lowest term in the angular momentum expansion of the amplitude, but for all orders in the partial wave expansion. In particular, we carefully account for the effects of the spin of the annihilating particle on the symmetry of the two-particle wave function. This work focuses on strongly interacting particles of arbitrary spin in the triplet, sextet and octet color representations. For typical velocities during freeze-out, we find that including Sommerfeld corrections on the next-to-leading order partial wave leads to modifications of up to ten to twenty percent on the total annihilation cross section. Complementary to QCD, we generalize our results to particles charged under an arbitrary unbroken $SU(N)$ gauge group, as encountered in dark glueball models. In connection with this paper a \texttt{Mathematica} notebook is provided to compute the Sommerfeld corrections for colored particles up to arbitrary order in the angular momentum expansion.}

\begin{document}

\maketitle
\clearpage

\section{Introduction}
\label{sec:introduction}
Sommerfeld corrections~\cite{Sommerfeld:1931aa} through long range interactions play a critical role in a plethora of thermal dark matter scenarios. Affecting primarily particles with low velocity, they have been shown, for instance, to tremendously enhance the dark matter annihilation rate at the galactic center in various models~\cite{Hisano:2004ds,ArkaniHamed:2008qn,Lattanzi:2008qa}. In particular, the predicted enhancement of the annihilation rate in the galactic center for pure wino dark matter has allowed to strongly restrict this supersymmetric scenario~\cite{Cohen:2013ama,Fan:2013faa,Hryczuk:2014hpa,Beneke:2016jpw}. For multi-TeV weakly interacting dark matter, Sommerfeld corrections also typically lead to order one modifications of its relic density~\cite{Hisano:2006nn,Hryczuk:2010zi,Beneke:2014gja,Beneke:2016ync}, often significantly weakening the upper bound on the dark matter mass derived from the Planck measurement~\cite{Ade:2015xua}.

Even in the sub-TeV regime, Sommerfeld corrections become significant in models involving long range interactions with order one couplings. In particular, a wide range of dark matter models --- such as supersymmetry or simplified models of coannihilation~\cite{Baker:2015qna,Buschmann:2016hkc} --- involve strongly interacting particles in the dark sector. Although the strong interaction is short ranged at low energies, in the early universe the non-relativistic QCD potential can be approximated by a Coulomb potential at tree-level~\cite{Strumia:2008cf,deSimone:2014pda}. Strongly interacting dark sector particles would therefore experience sizable long-range interactions through gluon exchange. These interactions would in turn significantly affect the annihilation rate into quarks and gluons for masses as low as $\mathcal{O}(100$~GeV$)$. Computing this rate accurately is crucial in various scenarios, such as in models where colored particles can survive until short before BBN, or models where dark matter coannihilates with a colored partner. In the latter case, the dark matter depletion will in fact be driven by the annihilation of its coannihilation partner through strong interaction in most of the parameter space.

Analytical and numerical computations of the Sommerfeld modified annihilation rate for heavy colored particles have been carried out in various studies~\cite{Ellis:2015vaa,Ibarra:2015nca,Strumia:2008cf,deSimone:2014pda,Harz:2014gaa}. Notably, reference~\cite{deSimone:2014pda} introduces a general method to decompose the QCD potential into a sum of Coulomb potentials for different possible $SU(3)$ representations of the colored dark sector particle. However, all the existing results only correctly describe corrections to the $s$-wave cross sections, while higher order effects are significant. For uncolored particles, the Sommerfeld effect has been computed beyond the $s$-wave in~\cite{Cassel:2009wt,Cassel:2009pu,Iengo:2009ni}. Yet, these results are consistent only when the annihilation amplitude is dominated by a single angular momentum component --- typically $s$-wave or $p$-wave. In addition, extending the aforementioned results to colored particles is non-trivial.

Aside from $SU(3)$, Sommerfeld corrections for dark sector particles charged under a general $SU(N)$ gauge group have not been considered in the literature. These non-perturbative effects related to the dark gauge interaction can significantly modify the annihilation cross-sections of new charged particles before their freeze-out or, in the case of $SU(N)$ relics, impact the predicted indirect detection signal. In face of growing interest in the cosmological role of new gauge groups~\cite{Gherghetta:2016bcc,Buen-Abad:2015ova,Ko:2016fcd,Feng:2011ik,Boddy:2014qxa,Boddy:2014yra,Harigaya:2016nlg,Soni:2017nlm}, the Sommerfeld effect should be derived and implemented also in the case of non-SM interactions.

In this paper, we present a robust and general framework to analytically compute the Sommerfeld corrections for the annihilation of dark sector particles charged either under QCD or $SU(N)$. Instead of considering only the leading term in the angular momentum expansion of the amplitude, our approach operates on its complete partial wave expansion into initial states of definite orbital angular momentum $l$ and spin $s$. Our study focuses on extensions of the Standard Model with a SM singlet dark matter candidate and one heavy new particle $\Phi$ which can be either a scalar, a fermion or a vector. We  first consider scenarios where $\Phi$ is a triplet, sextet or octet of $SU(3)$ and annihilates into quark and gluon pairs. We then generalize these results to the case where $\Phi$ is charged under either the fundamental or the adjoint representation of a dark $SU(N)$ gauge group. We discuss direct applications of these new results in glueball dark matter scenarios. In a companion paper~\cite{ElHedri:2017nny}, we perform a general study of the relic density and collider constraints on dark matter models with a colored coannihilation partner. In these scenarios, the annihilation of $\Phi$ through strong interactions drives the dark matter depletion and the derived constraints on the models do not depend on new physics couplings.

The work is organized as follows. In section~\ref{sec:sommerfeld:partialwaves} we discuss the analytic derivation of Sommerfeld corrections to annihilation processes for arbitrary partial waves and with any momentum dependence. In section~\ref{sec:sommerfeld:qcd} we review Sommerfeld corrections for QCD potentials in a manner that is applicable to annihilation of particles with arbitrary color representation. The approaches in sections~\ref{sec:sommerfeld:partialwaves} and~\ref{sec:sommerfeld:qcd} are orthogonal and can be combined into a general prescription for the annihilation of colored particles. In section~\ref{sec:colored:dark:sector} we show that these Sommerfeld effects are significant for colored dark sectors. In addition to QCD we discuss the Sommerfeld correction for dark sectors charged under $SU(N)$ in section~\ref{sec:sun:dark:sector}. We conclude in section~\ref{sec:conclusions} and discuss more exotic colored dark sectors in appendix~\ref{sec:color:decomposition}.

\section{Sommerfeld corrections for partial waves}
\label{sec:sommerfeld:partialwaves}
Accurately computing the Sommerfeld corrections for an arbitrary process can prove a daunting task that often has to be performed numerically. Annihilations in the dark sector, however, involve heavy particles and can therefore be studied in the non-relativistic limit. In this limit, the tree-level amplitude for a given process can be reliably approximated by a partial wave expansion in the orbital angular momentum $l$ and the spin $s$ in either the initial or final state. Notably, for a $2\rightarrow 2$ process with two scalar fields in the initial state, this expansion would be of the form
\begin{equation}
	\mathcal{M}(p, \theta, \phi) = \sum_{l,m} F_{lm}(p) Y_{lm} (\theta,\phi) , 
\end{equation}
where $p$ is the magnitude of the initial state momentum in the center-of-mass frame, $(\theta,\phi)$ are the scattering angles and $Y_{lm}(\theta, \phi)$ are the spherical harmonics. Without loss of generality, the radial part of the amplitude can be expanded in powers of $p$ such that the lowest-order contribution for a given $l$ is $p^l$
\begin{equation} \label{eq:fexpansion}
	F_{lm}(p) = \sum_{n\geq 0} \alpha_{lmn} \, p^{l+2n} .
\end{equation}

For amplitudes that are dominated by a single partial wave process, the Sommerfeld corrections can be expressed as an overall multiplicative factor to the tree-level cross section
\begin{equation}
	\sigma_\mathrm{Sommerfeld} = S \sigma_\mathrm{perturbative} . 
\end{equation}
The rescaling factor $S$ encodes the modification of the transition amplitude by a distorting potential $V$ (modeling the long-range interactions in the non-relativistic limit) acting on the initial particle wave functions. 
For a Coulomb potential $V = -A / r$ in particular, this factor has a simple analytic form in the $s$-wave~\cite{Sommerfeld:1931aa}
\begin{equation} \label{eq:sommerfeld:corrections:lowestorder}
	S(x) = \frac{\pi x}{1 - e^{-\pi x}}, \quad x = \frac{A}{\beta} .
\end{equation}
where $\beta$ is the velocity of the incoming particles in the center-of-mass frame. Positive $A$ corresponds to an attractive potential which leads to an enhancement of the perturbative result, while negative $A$ results in a depletion of the cross section due to the repulsive interaction. Analytical formulae for the Sommerfeld correction factors for higher waves have been computed in~\cite{Cassel:2009wt,Iengo:2009ni} assuming the amplitude is proportional to $p^l$ for the $l$th partial wave. This has been extended upon slightly in~\cite{Cassel:2009pu} allowing for a single term with a momentum dependence of $p^{l+2n}$ with $n \geq 0$. Here, we extend these results to a full  expansion of the annihilation amplitude into orbital angular momentum and spin states $(l, s)$ up to an arbitrary $l_{\text{max}}$. In particular, we allow the different terms of the expansion to coexist and we take higher order terms in equation~\eqref{eq:fexpansion} into account. In the rest of this section we consider a Coulomb potential and do not make assumptions about the spin of the initial state particles.

\subsection{Partial wave expansion}
\label{sec:partial:wave:expansion}
For a given field $\Phi$, the $\Phi \, \overline{\Phi} \to \textrm{SM} \, \textrm{SM}$ amplitude can be expanded into orbital angular momentum and spin states $(l, s)$. The reasons for doing this expansion are manifold. First, as argued at the beginning of this section, this expansion can be interpreted as a velocity expansion, which would provide an accurate approximation of the annihilation amplitude for non-relativistic particles. Moreover, as we will explain in section~\ref{sec:partial:wave:sommerfeld}, obtaining Sommerfeld corrections involves computing the non-relativistic wave function for the two $\Phi$ scattering states. In our case, this wave function is a solution of the Schr\"odinger equation for a Coulomb potential. As shown in~\cite{Cassel:2009pu}, expanding both the scattering state wave function and the annihilation amplitude considerably simplifies calculations. This leads to a set of independent equations for each partial wave and allows to obtain analytical formulae for the Sommerfeld corrected matrix element $\mathcal{M}^{(S)}_{ls}$. Note that, since the different $(l, s)$ states are orthogonal to each other, the final cross section will be of the form
\begin{equation}
	\sigma^{(S)} \propto \sum_{l,s} \left| \mathcal{M}^{(S)}_{ls} \right|^2 .
\end{equation}

Another notable advantage of using a $(l, s)$ decomposition is that for identical particles the overall form of a given $(l, s)$ state is strongly constrained by CP conservation. For particles carrying no other quantum numbers than the ones associated to the Lorentz group, a CP transformation multiplies the initial or final state wave function by $(-1)^{l+s}$. Only states with even $l + s$ would therefore have a non-zero amplitude. For colored particles, on the other hand, the color factor in the amplitude can be decomposed into two parts, respectively symmetric and antisymmetric under particle exchange. States with even $l+s$ will be proportional to the symmetric part while states with odd $l+s$ will be proportional to the antisymmetric part. For $\Phi \, \overline{\Phi} \to g^a \, g^b$ in particular, since the gluons are identical particles, the contributions from states with even $l+s$ will be proportional to $\left\{ T^a_{\irrep{R}}, T^b_{\irrep{R}} \right\}$ while the ones for states with odd $l+s$ will be proportional to $\left[ T^a_{\irrep{R}}, T^b_{\irrep{R}} \right]$ where $T^a_{\irrep{R}}$ is the generator for the representation $\irrep{R}$ of $\Phi$. This color factor dependence will allow us to introduce a generic procedure to decompose the amplitude into definite color states as we will describe in section~\ref{sec:sommerfeld:qcd}. The same arguments apply to the case of $SU(N)$, which will be discussed in section~\ref{sec:sun:dark:sector}.

In what follows, we consider a $2 \rightarrow 2$ annihilation process in the center-of-mass frame and in the spin basis. Without loss of generality we choose the final state particles to be along the $\hat{\mathbf{z}}$-axis and denote the scattering angles in the initial state by $(\theta, \phi)$. With $\left\{m_1, m_2\right\}$ and $\left\{m_3, m_4\right\}$ being the individual spin projections on the $\hat{\mathbf{z}}$-axis in the initial and final states respectively, the total annihilation amplitude is defined as an element of the transition matrix $T$
\begin{equation} \label{eq:amplitude}
	\mathcal{T}_{fi} \left( p, \theta, \phi \right) = \langle p_f; 0 0; m_3 m_4 | T | p; \theta \phi; m_1 m_2 \rangle .
\end{equation}
Here, $p$ and $p_f$ being the magnitudes of the momenta in the initial and final states respectively. The information about the total spins $s_{1,2,3,4}$ in the initial and final state is omitted here for compactness of notation. Further details about the computation of the total amplitude --- notably our definitions for the momenta and the polarization vectors --- are provided in appendix~\ref{sec:xsec:partial:wave}.

Decomposing the initial state into states of definite orbital angular momentum $(l, l_z)$, the amplitude can be rewritten as\footnote{The spherical harmonics $Y_l^m(\theta, \phi)$ are normalized as \begin{equation*} \int Y_l^m(\theta, \phi) Y_{l'}^{m'}(\theta, \phi) \,\mathrm{d}\Omega = \delta_{ll'}\delta_{mm'} .\end{equation*}}
\begin{equation}
	\mathcal{T}_{fi} \left( p, \theta, \phi \right) = \sum_{l, l_z} \langle p_f; 0 0; m_3 m_4 | T | p; l l_z; m_1 m_2 \rangle \, Y^{l_z}_l(\theta, \phi) .
\end{equation}
A given $| p_i; l l_z; m_1 m_2 \rangle$ state can be decomposed into $| p; l l_z; s s_z \rangle$ states using Clebsch-Gordan coefficients
\begin{equation}
	| p; l l_z; m_1 m_2 \rangle = \sum_{s, s_z} \langle s_1 m_1 s_2 m_2 | s s_z \rangle | p; l l_z; s s_z \rangle ,
\end{equation}
where $s_1, s_2$ are the total spins of the incoming particles. The total amplitude can then be written as
\begin{equation} \label{eq:amplitude:ls:decomp}
	\mathcal{T}_{fi} \left( p, \theta, \phi \right) = \sum_{l, l_z} \sum_{s, s_z} Y^{l_z}_l(\theta, \phi) \langle s_1 m_1 s_2 m_2 | s s_z \rangle \mathcal{M}(p; l l_z; s s_z; m_3 m_4) .
\end{equation}
The matrix element $\mathcal{M}(p; l l_z; s s_z; m_3 m_4)$ corresponds to the contribution of a single initial state $| p_i; l l_z; s s_z \rangle$ to the total amplitude. Knowing $\mathcal{T}_{fi}$, this matrix element can be computed using
\begin{equation}
	\begin{aligned}
		\mathcal{M}(p; l l_z; s s_z; m_3 m_4) & \equiv \langle p_f; 0 0; m_3 m_4 | T | p; l l_z; s m_s \rangle \\
		& = \sum_{m_1, m_2} \langle s_1 m_1 s_2 m_2 | s s_z \rangle \int \mathrm{d} \Omega\, Y^{l_z \, *}_l(\theta, \phi) \mathcal{T}_{fi} \left( p, \theta, \phi \right) .
	\end{aligned}
\end{equation}

Since the $(l,s)$ components of the amplitude are orthogonal the total cross section is of the form
\begin{equation} \label{eq:total:cross:section:partial:waves}
	\sigma = \frac{1}{64 \pi^2 s} \frac{1}{\sqrt{1 - \frac{4m_\Phi^2}{s}}} \dfrac{1}{d_\Phi^2 d_{\irrep{R}}^2} \sum_{m_3, m_4} \sum_{l, l_z} \sum_{s, s_z} \left| \mathcal{M}(p; l l_z; s s_z; m_3 m_4) \right|^2 ,
\end{equation}
where $d_\Phi$ is the number of degrees of freedom of the field $\Phi$ and $d_{\irrep{R}}$ is the dimensionality of the color representation of $\Phi$. Another factor $\frac{1}{2}$ needs to be included for identical final state particles like two gluons.

As mentioned at the beginning of this section, the amplitude for a given $l$ can be expanded in powers of the magnitude of the incoming particle momentum $p = \sqrt{\frac{s}{4} - m_\Phi^2}$, with the lowest-order contribution for a given $l$ being $\mathcal{O}(p^l)$. We can therefore write
\begin{equation} \label{eq:lsdecomp}
	\mathcal{M}(p; l l_z; s s_z; m_3 m_4) = \sum_{n\ge 0} \alpha^{(m_3,m_4)}_{ll_zss_z,n} \, p^{l+2n} .
\end{equation}
Since the matrix element is now expanded in the momentum and in $l$ we can apply the Sommerfeld corrections to each of the terms in equation~\eqref{eq:lsdecomp}. This will be derived in the next section and the total Sommerfeld-corrected cross sections can then be obtained by the use of equation~\eqref{eq:total:cross:section:partial:waves}.

\subsection{Sommerfeld corrections}
\label{sec:partial:wave:sommerfeld}
The Sommerfeld effect is a non-perturbative phenomenon caused by the distortion of the scattering amplitude of two particles through long range interactions. This distortion occurs primarily at low velocities and therefore can particularly affect non-relativistic particles such as the ones in the dark sector. 

\begin{figure}[!t]
	\centering
	\begin{tikzpicture}[line width=1.4pt, scale=1]
	\draw[fermionbar] (-2.8,0.8)--(-3.6,0.8);
	\draw[fermion] (-2.8,-0.8)--(-3.6,-0.8);
	\draw[fermionbar] (-2.2,0.8)--(-2.8,0.8);
	\draw[fermion] (-2.2,-0.8)--(-2.8,-0.8);
	\draw[fermionbar] (-1.2,0.8)--(-2.2,0.8);
	\draw[fermion] (-1.2,-0.8)--(-2.2,-0.8);
	\draw[fermionna] (-1.2,0.8)--(-0.8,0.8);
	\draw[fermionna] (-1.2,-0.8)--(-0.8,-0.8);
	\draw[gluon] (-2.8,0.8)--(-2.8,-0.8);	
	\draw[gluon] (-2.2,0.8)--(-2.2,-0.8);
	\draw[gluon] (-1.2,0.8)--(-1.2,-0.8);
	\draw[fermionna] (0.8,0.8)--(0,0);
	\draw[fermionna] (0.8,-0.8)--(0,0);
	\draw[fermion] (-0.8,0.8)--(0,0);
	\draw[fermionbar] (-0.8,-0.8)--(0,0);
	\draw[fill=black] (0,0) circle (3.0mm);
	\draw[fill=white] (0,0) circle (2.9mm);
	\begin{scope}
		\clip (0,0) circle (3.0mm);
		\foreach \x in {-.9,-.75,...,.3}
		\draw[line width=1 pt] (\x,-.3) -- (\x+.6,.3);
	\end{scope}
	
	\draw[fill=black] (-1.5,0) circle (0.1mm);
	\draw[fill=black] (-1.7,0) circle (0.1mm);
	\draw[fill=black] (-1.9,0) circle (0.1mm);
	
	\node at (-3.8,0.8) {$\Phi$};
	\node at (-3.1,0.0) {$g$};
	\node at (-3.8,-0.8) {$\overline{\Phi}$};
	\node at (1.15,0.8) {SM};
	\node at (1.15,-0.8) {SM};
\end{tikzpicture}
	\caption{Sommerfeld ladder diagram for the annihilation of $\Phi$ into Standard Model particles.}
	\label{fig:sommerfeld_ladder}
\end{figure}
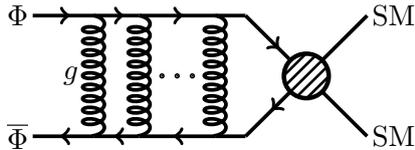

Although non-perturbative, the Sommerfeld effect can be approximately modeled by considering the limit of Feynman diagrams with an infinite number of particle exchanges~\cite{Cassel:2009pu}. These diagrams should in general include all the possible two-particle irreducible interactions, which would make the computation of the final amplitude particularly cumbersome. For non-relativistic particles, however, the final amplitude is dominated by ladder diagrams with an infinite number of one-particle exchange iterations such as the one shown in figure~\ref{fig:sommerfeld_ladder}. For a given $2\rightarrow n$ process with the Sommerfeld effect occurring in the initial state, the amplitude then verifies the following recursion relation~\cite{Visinelli:2010vg}
\begin{equation}
	\mathcal{M}^{(S)}_{\beta\alpha} = \mathcal{M}^0_{\beta\alpha} + \int d\gamma \frac{\mathcal{M}^{(S)}_{\beta\gamma} V_{\gamma\alpha}}{E_\alpha - E_\gamma + i \epsilon} ,
\end{equation}
where $\alpha$ and $\beta$ are the initial and final states respectively and the integral over $\gamma$ represents the sum over all possible intermediate states. $\mathcal{M}^0_{\beta\alpha}$ is the perturbative scattering amplitude corresponding to the exchange of one particle and $V_{\gamma\alpha}$ is the non-relativistic interaction potential distorting the initial state $\alpha$.

The interaction potential $V_{\gamma\alpha}$ can be rewritten as
\begin{equation}
	V_{\gamma \alpha} = \langle k; \theta_{k} \phi_{k}; m_1 m_2 | \hat{V} | p; \theta_p \phi_p; m_a m_b \rangle ,
\end{equation}
where $\left\{m_1, m_2\right\}$ and $\left\{m_a, m_b\right\}$ are the $z$-components of the spins of the $\alpha$ and $\gamma$ states respectively and $p, k = |\mathbf{p}|, |\mathbf{k}|$ are the magnitudes of the momenta $\mathbf{p}$ and $\mathbf{k}$ in these states. In the rest of this work, we will focus on a spin-independent spherically symmetric potential $V(|\mathbf{p}-\mathbf{k}|)$. We can therefore factor out the spin states, which gives 
\begin{equation}
	\begin{aligned}
		V_{\gamma\alpha} & = \langle m_1 m_2 | m_a m_b \rangle \langle k; \theta_k \phi_k | \hat{V} | p; \theta_p \phi_p \rangle \\
		& = \delta_{m_1 m_a} \delta_{m_2 m_b} V(|\mathbf{p} - \mathbf{k}|) .
	\end{aligned}
\end{equation}

For initial and final state spins $\textbf{m}_i = \left\{ m_1, m_2 \right\}$ and $\textbf{m}_f = \left\{ m_3, m_4 \right\}$, the Sommerfeld-corrected amplitude can be expressed as
\begin{equation}
	\mathcal{M}^{(S)}_{\textbf{m}_f \textbf{m}_i}(\mathbf{p}) = \mathcal{M}^0_{\textbf{m}_f \textbf{m}_i}(\mathbf{p}) + \int \frac{\mathrm{d}^3k}{(2 \pi)^3} \frac{\mathcal{M}^{(S)}_{\textbf{m}_f \textbf{m}_i}(\mathbf{k}) V(|\mathbf{p} - \mathbf{k}|)}{E_\alpha - E_\gamma + i \epsilon} . 
\end{equation}
In the non-relativistic limit, the denominator can be rewritten as
\begin{equation}
	(E_\alpha - E_\gamma)^{-1} \approx \mathcal{E} - \frac{\mathbf{k}^2}{2 \mu} ,
\end{equation}
where $\mathcal{E}$ is the total energy of the system and $\mu$ is its reduced mass. For two particles of identical mass $m$, $\mu = \frac{m}{2}$. If the initial states are off-shell, that is $\frac{\mathbf{\widetilde{p}}^2}{2\mu} \neq \mathcal{E}$, we can define~\cite{Iengo:2009ni}
\begin{equation}
	\Phi_{\textbf{m}_f \textbf{m}_i}(\widetilde{\mathbf{p}}) = \frac{\mathcal{M}^{(S)}_{\textbf{m}_f \textbf{m}_i}(\widetilde{\mathbf{p}})}{\frac{\widetilde{\mathbf{p}}^2}{2 \mu} - \mathcal{E}} ,
\end{equation}
which verifies
\begin{equation}
	\left(\frac{\widetilde{\mathbf{p}}^2}{2 \mu} -\mathcal{E}\right) \Phi_{\textbf{m}_f \textbf{m}_i} (\widetilde{\mathbf{p}}) = \mathcal{M}^0_{\textbf{m}_f \textbf{m}_i} (\widetilde{\mathbf{p}}) - \int \frac{\mathrm{d}^3 k}{(2 \pi)^3} \Phi_{\textbf{m}_f \textbf{m}_i}(\mathbf{k}) V(|\widetilde{\mathbf{p}} - \mathbf{k}|) .
\end{equation}
In position space (we go from $\widetilde{\mathbf{p}}$ to $\mathbf{r}$) this becomes 
\begin{equation}
	\left( \frac{-\mathbf{\nabla}^2}{2 \mu} + V(r) - \mathcal{E} \right) \widetilde{\Phi}_{\textbf{m}_f \textbf{m}_i}(\mathbf{r}) = U^0_{\textbf{m}_f \textbf{m}_i}(\mathbf{r}) ,
\end{equation}
where
\begin{equation}
	U^0_{\textbf{m}_f \textbf{m}_i}(\mathbf{r}) = \int \frac{d^3 q}{(2 \pi)^3} e^{i \mathbf{r} \cdot \mathbf{q}} \mathcal{M}^0_{\textbf{m}_f \textbf{m}_i} (\mathbf{q}) . 
\end{equation}
The final amplitude can now be computed by putting the initial states back on-shell
\begin{equation}
	\mathcal{M}^{(S)}_{\textbf{m}_f \textbf{m}_i}(\mathbf{p}) = \mathrm{lim}_{\widetilde{\mathbf{p}}\rightarrow \mathbf{p}} \left(\frac{\widetilde{\mathbf{p}}^2}{2 \mu}- \mathcal{E} \right) \Phi_{\textbf{m}_f \textbf{m}_i}(\widetilde{\mathbf{p}}) \quad \quad \mathrm{with} \quad \frac{\mathbf{p}^2}{2 \mu} = \mathcal{E} ,
\end{equation}
which leads to~\cite{Iengo:2009ni,Cassel:2009wt}
\begin{equation} \label{eq:nonpert}
	\mathcal{M}^{(S)}_{\textbf{m}_f \textbf{m}_i}(\mathbf{p}) = \int \frac{\mathrm{d}^3 q}{(2 \pi)^3} \, \mathcal{M}^0_{\textbf{m}_f \textbf{m}_i}(\mathbf{q}) \, \phi_{\mathbf{p}}(\mathbf{q}) ,
\end{equation}
where $\phi_{\mathbf{p}}(\mathbf{q})$ obeys the traditional Schr\"odinger equation in position space
\begin{equation}
	\left(\frac{-\nabla^2}{2\mu} + V(r) - \frac{p^2}{2 \mu} \right) \widetilde{\phi}_{\mathbf{p}}(\mathbf{r}) = 0 \quad \quad \mathrm{with} \quad \frac{p^2}{2 \mu} \equiv \mathcal{E}. 
\end{equation}

For a potential of the form $V(|\mathbf{p} - \mathbf{q}|)$, the wave function can be rewritten as
\begin{equation}
	\phi_{\mathbf{p}}(\mathbf{q}) = \phi(p, q, \hat{\mathbf{p}} \cdot \hat{\mathbf{q}}) ,
\end{equation}
and can therefore be expanded in Legendre polynomials and in spherical harmonics
\begin{equation} \label{eq:wavefunction:decomp}
	\begin{aligned}
		\phi_{\mathbf{p}}(\mathbf{q}) & = \sum_l \frac{2l+1}{4\pi} \, F_l (p, q) P_l (\hat{\mathbf{p}} \cdot \hat{\mathbf{q}}) \\
		& = \sum_{l, l_z} F_l (p, q) Y_{l}^{l_z \, *}(\theta_q, \phi_q) Y_l^{l_z}(\theta_p, \phi_p) .
	\end{aligned}
\end{equation}
As shown in equation~\eqref{eq:amplitude:ls:decomp}, the perturbative amplitude can be expanded in spherical harmonics as well
\begin{equation} \label{eq:amplitude:pert:decomp}
	\mathcal{M}^0_{\textbf{m}_i \textbf{m}_f}(\mathbf{q}) = \sum_{l,l_z} \sum_{s, s_z} \langle s_1 m_1 s_2 m_2 | s s_z \rangle \mathcal{M}(q; l l_z; s s_z; \textbf{m}_f) \, Y^{l_z}_l(\theta_q, \phi_q) .
\end{equation}
Injecting equations~\eqref{eq:wavefunction:decomp} and \eqref{eq:amplitude:pert:decomp} into equation~\eqref{eq:nonpert}, the Sommerfeld corrected matrix element can then be decomposed as
\begin{equation} \label{eq:nonpert:decomp}
	\begin{aligned}
		\mathcal{M}^{(S)}_{\textbf{m}_i\textbf{m}_f}(\mathbf{p}) & = \sum_{l, l_z} \sum_{s, s_z} \sum_{l', l_z'} \int \frac{\mathrm{d}^3 q}{(2\pi)^3} \, \langle s_1 m_1 s_2 m_2 | s s_z \rangle \mathcal{M}(q; l l_z; s s_z; \textbf{m}_f) \, F_{l'} (p, q) \\
		& \qquad \qquad \qquad \quad \times Y^{l_z}_l(\theta_q, \phi_q) Y^{l_z'}_{l'}(\theta_q, \phi_q) Y^{l_z'}_{l'}(\theta_p, \phi_p) \\
		& = \sum_{l, l_z} \sum_{s, s_z} \langle s_1 m_1 s_2 m_2 | s s_z \rangle \! \int \! \frac{q^2\mathrm{d} q}{2\pi^2}\, \mathcal{M}(q; l l_z; s s_z; \textbf{m}_f) \, F_l (p, q)\, Y^{l_z}_{l}(\theta_p, \phi_p) .
	\end{aligned}
\end{equation}
Here, in the last line we used the orthogonality relations for the spherical harmonics. The Sommerfeld corrected amplitude for a given $(l,l_z,s,s_z)$ state then takes the following simple form
\begin{equation}
	\mathcal{M}^{(S)}_{ll_z;ss_z;\textbf{m}_f}(p) = \int \frac{q^2\mathrm{d} q}{2\pi^2}\, \mathcal{M}(q; l l_z; s s_z; \textbf{m}_f) \, F_l (p, q) .
\end{equation}
Using equation~\eqref{eq:lsdecomp}, we can re-express this amplitude as
\begin{equation}
	\mathcal{M}^{(S)}_{ll_z;ss_z;\textbf{m}_f}(p) = \sum_{n\ge 0} \alpha^{\textbf{m}_f}_{ll_z;ss_z;n} \int \frac{\mathrm{d} q}{2\pi^2}\, q^{l + 2n + 2} \, F_l (p, q) .
\end{equation}
As shown in~\cite{Cassel:2009pu}, the integrals can be rewritten as functions of the derivatives of the radial components of the wave function $R_{pl}(r)$
\begin{equation}
	\int \frac{\mathrm{d} q}{2 \pi^2}\, q^{l + 2n + 2} \, F_l (p, q) = \frac{2^n\, n!(2l + 2n + 1)!!}{(-1)^n (-i)^l (l+2n)!} \, \left.\frac{\partial^{l + 2n} R_{pl}(r)}{\partial r^{l+2n}} \right|_{r = 0} .
\end{equation}
For a Coulomb potential $V = -A / r$, the radial components of the wave function can be computed analytically and are equal to
\begin{equation} \label{eq:coulomb:radial:wavefunction}
	R_{pl}(z; x) = e^\frac{\pi x}{4} e^\frac{-iz}{2} z^l \sum_{j = 0}^\infty \frac{\Gamma \left( 1 + \frac{i x}{2} + l + j \right)}{(2l + 1 + j)!}\frac{(iz)^j}{j!} ,
\end{equation}
where $z = 2 r p$, $x = A m / p$ and with $p$ and $m$ the momentum and mass of the incoming particles. In our study, since we consider strong interactions, $A$ will be proportional to the QCD coupling $\alpha_s$ or the $SU(N)$ coupling $\alpha_N$.

Using the expression given in equation~\eqref{eq:coulomb:radial:wavefunction}, we can then write
\begin{equation} \label{eq:sommerfeld:corrected:amplitude}
	\mathcal{M}^{(S)}_{ll_z;ss_z;\textbf{m}_f}(p) = \sum_{n\ge 0} \alpha^{\textbf{m}_f}_{ll_z;ss_z;n} p^{l + 2n} \mathcal{C}_{l} (x) \mathcal{D}_{ln} (x).
\end{equation}
The Sommerfeld factors $\mathcal{C}_{l} (x)$ and $\mathcal{D}_{ln} (x)$ are given by
\begin{equation} \label{eq:cl:dln}
	\begin{aligned}
		\mathcal{C}_l (x) & = \frac{1}{(-i)^l} \, e^\frac{\pi x}{4} \,\Gamma\left(1 + \frac{i x}{2}\right) \prod_{b = 1}^{l} \left( 1 + \frac{i x}{2 b} \right) \\
		\mathcal{D}_{ln} (x) & = \frac{n! (2l+2n+1)!}{(l+n)!} \sum_{j = 0}^{2n} \frac{(-2)^j (l+j)!}{j! (2n-j)! (2l+j+1)!} \left[\prod_{b = l+1}^{l+j}\left(1 + \frac{i x}{2b}\right)\right] ,
	\end{aligned}
\end{equation} 
where $\mathcal{C}_l (x)$ is the correction to the amplitude for a perturbative matrix element of the form $p^l Y_l^{l_z}(\theta_p, \phi_p)$. Note here that $\mathcal{D}_{l0} (x) = 1$ by construction. The Sommerfeld-corrected squared matrix element for an $(l, l_z, s, s_z)$ initial state as given in equation~\eqref{eq:total:cross:section:partial:waves} can then be written as
\begin{equation} \label{eq:sommerfeld:enhanced:sqmatrix}
	\left| \mathcal{M}^{(S)}_{ll_z;ss_z;\textbf{m}_f}(p) \right|^2 = S_l (x) \sum_{n,n'} \alpha^{\textbf{m}_f}_{ll_z;ss_z;n} \left(\alpha^{\textbf{m}_f}_{ll_z;ss_z;n'}\right)^* \mathcal{D}_{ln} (x) \mathcal{D}_{ln'}^* (x) \, p^{2(l + n + n')} ,
\end{equation}
where
\begin{equation} 
	S_{l} (x) = |\mathcal{C}_{l} (x)|^2 = \frac{\pi x}{1-e^{-\pi x}} \prod_{b = 1}^{l} \left( 1 + \frac{x^2}{4 b^2} \right)
\end{equation}
is the Sommerfeld correction for a Coulomb potential and for a perturbative amplitude of the form $p^l Y_{l}^{l_z}(\theta_p, \phi_p)$~\cite{Cassel:2009wt,Iengo:2009ni}. Here, we used $\left|\Gamma(1 + i b)\right| = \sqrt{\pi b \, \mathrm{csch}(\pi b)}$. Note that, in equation~\eqref{eq:sommerfeld:enhanced:sqmatrix}, since higher order terms are taken into account in the momentum expansion of the perturbative amplitude, the Sommerfeld corrections can no longer be factored out. The total Sommerfeld-corrected cross section is then obtained by plugging equation~\eqref{eq:sommerfeld:enhanced:sqmatrix} into equation~\eqref{eq:total:cross:section:partial:waves}.

\subsection{Convergence and strategy}
\label{sec:partial:wave:strategy}
The Sommerfeld corrections as given in equation~\eqref{eq:sommerfeld:enhanced:sqmatrix} depend on $l$, $n$ and on inverse powers of the velocity through $x$. In the perturbative regime, the angular momentum expansion and velocity expansion of the cross section are closely related. For a given angular momentum $l$, the lowest-order term of the perturbative amplitude is at best $\mathcal{O}(v^l)$ or equivalently $\mathcal{O}(p^l)$. This relation is however lost when incorporating the Sommerfeld corrections. As shown in equation~\eqref{eq:cl:dln}, at low velocity, the Sommerfeld factor for a given $(l, n)$ is $\mathcal{O}(p^{-l - 2n - \frac{1}{2}})$. For a momentum expansion of the perturbative amplitude of the form
\begin{equation}
	\mathcal{M}_{ll_z;ss_z;\textbf{m}_f}^0(p) = \sum_{n\ge 0} \alpha_{ll_z;ss_z;n}^{\textbf{m}_f} p^{l+2n} ,
\end{equation}
the convergence in the momentum is then jeopardized by the Sommerfeld factor. The lowest-order term of the momentum expansion of the Sommerfeld-corrected amplitude given in equation~\eqref{eq:sommerfeld:corrected:amplitude} becomes
\begin{equation} \label{eq:sommerfeld:convergence}
	\begin{aligned}
		\mathcal{M}_{ll_z;ss_z;\textbf{m}_f}^{(S)}(p) & = \sqrt{\frac{\pi A m}{p}}\sum_{n\ge 0} (-1)^{l+n} \alpha_{ll_z;ss_z;n}^{\textbf{m}_f} m^{l + 2n} \,\frac{A^{l+2n}}{2^l(l + n)!}\, \frac{n!}{(2n)!} + \mathcal{O}(p^\frac{1}{2}) \\
		& = \sqrt{\frac{\pi Am}{p}}\sum_{n\ge 0} \widetilde{\alpha}_{ll_z;ss_z;n}^{\textbf{m}_f} \,\frac{A^{l+2n}}{2^l(l + n)!}\, \frac{n!}{(2n)!} + \mathcal{O}(p^\frac{1}{2}) ,
	\end{aligned}
\end{equation}
where $\widetilde{\alpha}_{ll_z;ss_z;n}^{\textbf{m}_f} \equiv (-1)^{l+n} \alpha_{ll_z;ss_z;n}^{\textbf{m}_f} m^{l + 2n}$ is dimensionless. For any value of the orbital angular momentum $l$, the Sommerfeld-corrected amplitude can then contain terms of order $p^{-\frac{1}{2}}$. The convergence of the $(l, n)$ expansion of the cross section is now ensured by the factorial and $2^l$ terms as well as by the powers of $A$ since $A < 1$. Hence, the convergence is now in the orbital angular momentum $l$ instead of the velocity. Nonetheless, due to its factorial nature the convergence of the corrected cross section is at least as fast as the one of the perturbative cross section with $l$ and $n$. In fact, this non-trivial result ensures that the application of the Sommerfeld effect is a self-consistent procedure.

Since the angular momentum and velocity expansions of the Sommerfeld-corrected cross section are unrelated, we adopt the following strategy when calculating Sommerfeld corrections:
\begin{enumerate}
	\item Choose a maximal value $l_\mathrm{max}$ for the angular momentum expansion of both the perturbative and Sommerfeld-corrected cross sections. The choice for $l_\mathrm{max}$ determines the degree of precision for both expansions according to equations~\eqref{eq:lsdecomp} and~\eqref{eq:sommerfeld:convergence}. 
	\item For each value of $l$, include all expansion terms from equation~\eqref{eq:sommerfeld:enhanced:sqmatrix} with $n, n'$ satisfying $n + n' + l \leq l_\mathrm{max}$. This way, the highest order terms in this expansion are always $\mathcal{O}(p^{2l_{\mathrm{max}}-1})$. This requirement ensures the consistency of the expansion of the perturbative cross section in powers of the incoming momentum. 
	\item Finally, the total Sommerfeld-corrected cross section is obtained by injecting equation~\eqref{eq:sommerfeld:enhanced:sqmatrix} into equation~\eqref{eq:total:cross:section:partial:waves}.
\end{enumerate}
In this procedure the perturbative amplitude is fully expanded up to $p^{l_\mathrm{max}}$ and the perturbative cross section up to $p^{2 l_\mathrm{max} - 1}$. Applying Sommerfeld corrections to this expansion gives an angular momentum expansion of the final cross section up to $l_{\mathrm{max}}$. In sections~\ref{sec:sommerfeld:qcd} and~\ref{sec:sun:dark:sector} we describe how to embed non-Abelian gauge theories into this formalism. The results of applying this procedure to the annihilation of colored particles are shown in section~\ref{sec:sommerfeld:corrected:annihilation}.

\section{Sommerfeld corrections for QCD}
\label{sec:sommerfeld:qcd}
In the previous section we have computed analytic expressions for the Sommerfeld corrections of processes with arbitrary partial waves and momentum dependence. This derivation is based on a Coulomb potential, while the interactions between colored particles are governed by a QCD potential. An analytic prescription to decompose the QCD potential as a linear combination of Coulomb potentials has been first described in~\cite{Strumia:2008cf,deSimone:2014pda} for $s$-wave processes. In this section we extend this derivation to arbitrary partial waves and point out the differences to the leading order result. This extension allows for a treatment where higher order partial waves, arbitrary momentum dependence of the amplitude and QCD effects can all be taken into account. This prescription allows us to derive an analytic form for the Sommerfeld corrections of the annihilation of colored states which we apply to the colored dark sector in the next section.

\subsection{Decomposing the QCD potential}
\label{sec:decomposing:qcd:potential}
In order to analytically evaluate the Sommerfeld corrections through the exchange of soft gluons it is necessary to decompose the QCD potential into a set of Coulomb-like potentials. This is possible due to the fact the higher-order QCD potential takes the form~\cite{Fischler:1977yf,Schroder:1998vy,deSimone:2014pda}
\begin{equation} \label{eq:approximate:qcd:potential}
	V_\mathrm{QCD} = C \frac{\alpha_s (\hat{\mu})}{r} \left[ 1 + \frac{\alpha_s (\hat{\mu}) }{4 \pi} \left( c_1 + 2 c_2 (\gamma_E + \log \hat{\mu} r) \right) \right] \approx C \frac{\alpha_s (\hat{\mu} \approx 1/r)}{r} ,
\end{equation}
where $C$ is proportional to the quadratic Casimir. For example for the quark-antiquark potential $C = \frac{4}{3}$ and the one-loop coefficients are defined by $c_1 = \frac{31}{3} - \frac{10}{9} n_f$ and $c_2 = 11 - \frac{2}{3} n_f$, where $n_f$ is the number of active quark flavors at the scale $\hat{\mu}$. It shows that the QCD potential at higher orders can be approximated as a simple Coulomb-like form indicated on the right-hand side of equation~\eqref{eq:approximate:qcd:potential}. Now, as shown in~\cite{Strumia:2008cf,deSimone:2014pda}, the QCD potential between two particles of $SU(3)$ representations $\irrep{R}$ and $\irrep{R}'$ can be rewritten as a sum of Coulomb potentials of the form 
\begin{equation} \label{eq:potential:color:decomposition}
	V_{\irrep{R} \otimes \irrep{R}'} = \frac{\alpha_s (\hat{\mu})}{r} \sum_a T^a_{\irrep{R}} \otimes T^a_{\irrep{R}'} = \frac{\alpha_s (\hat{\mu})}{2r} \sum_{\irrep{Q}} \Big[ C_2(\irrep{Q}) \mathbbm{1}_{\irrep{Q}} - C_2(\irrep{R}) \mathbbm{1} - C_2(\irrep{R}') \mathbbm{1} \Big] ,
\end{equation}
where $\irrep{R} \otimes \irrep{R}' = \bigoplus_{\irrep{Q}} \irrep{Q}$ and $C_2(\irrep{R})$, $C_2(\irrep{R}')$ are the quadratic Casimir indices for $\irrep{R}$ and $\irrep{R}'$ respectively. Each irreducible $\irrep{Q}$ component of the initial-state wave function will then evolve independently in its respective potential. It is important to note here that $\alpha_s (\hat{\mu})$ must be evaluated at a much lower scale than the hard scale of the annihilation process, namely at scales similar to the momenta of the incoming particles. For clarity reasons we omit the scale dependence of $\alpha_s$ in the rest of this section. 

In what follows, we will consider particle-antiparticle annihilation with $\irrep{R} = \irrep{3}, \irrep{6}, \irrep{8}$ and $\irrep{R}'=\overline{\irrep{R}}$. The corresponding color decompositions are
\begin{equation} \label{eq:representation:product}
	\begin{aligned}
		\irrep{3} \otimes \overline{\irrep{3}} & = \irrep{1} \oplus \irrep{8} \\
		\irrep{6} \otimes \overline{\irrep{6}} & = \irrep{1} \oplus \irrep{8} \oplus \irrep{27} \\
		\irrep{8} \otimes \irrep{8} & = \irrep{1}_\textbf{S} \oplus \irrep{8}_\textbf{A} \oplus \irrep{8}_\textbf{S} \oplus \irrep{10}_\textbf{A} \oplus \overline{\irrep{10}}_\textbf{A} \oplus \irrep{27}_\textbf{S} .
	\end{aligned}
\end{equation}
The subscripts $\textbf{S}$ and $\textbf{A}$ indicate whether the representation is symmetric or antisymmetric respectively under the interchange of the two equal representations $\irrep{R}$ and $\irrep{R}'$. The quadratic Casimir indices ($C_2$) of these representations are given in table~\eqref{eq:casimir:invariants} along with the Dynkin indices, defined as $C(\irrep{R}) \delta^{ab} = \mathrm{tr} \! \left( T_{\irrep{R}}^a T_{\irrep{R}}^b \right)$.
\begin{equation} \label{eq:casimir:invariants}
	\begin{tabular}{c | c c c c c c c c}
		\irrep{R} & $\irrep{1}$ & $\irrep{3}$ & $\irrep{6}$ & $\irrep{8}$ & $\irrep{10}$ & $\irrep{15}$ & $\irrep{27}$ & $\irrep{64}$ \\
		\hline
		$C(\irrep{R})$ & $0$ & $\frac{1}{2}$ & $\frac{5}{2}$ & $3$ & $\frac{15}{2}$ & $10$ & $27$ & $120$ \\
		$C_2(\irrep{R})$ & $0$ & $\frac{4}{3}$ & $\frac{10}{3}$ & $3$ & $6$ & $\frac{16}{3}$ & $8$ & $15$ \\
	\end{tabular}
\end{equation}
Injecting equations~\eqref{eq:representation:product} and \eqref{eq:casimir:invariants} into equation~\eqref{eq:potential:color:decomposition}, we find
\begin{equation} \label{eq:decomposed:potentials}
	V_{\irrep{3} \otimes \overline{\irrep{3}}} = \! \frac{\alpha_s}{r} \! \left\{ \!\! \begin{aligned} - \frac{4}{3} & \quad (\irrep{1}) \\ + \frac{1}{6} & \quad (\irrep{8}) \end{aligned} \right. , \quad V_{\irrep{6} \otimes \overline{\irrep{6}}} = \! \frac{\alpha_s}{r} \! \left\{ \!\! \begin{aligned} - \frac{10}{3} & \quad (\irrep{1}) \\ - \frac{11}{6} & \quad (\irrep{8}) \\ + \frac{2}{3} & \quad (\irrep{27}) \end{aligned} \right. , \quad V_{\irrep{8} \otimes \irrep{8}} = \! \frac{\alpha_s}{r} \! \left\{ \!\! \begin{aligned} - 3 & \quad (\irrep{1}_\textbf{S}) \\ - \frac{3}{2} & \quad (\irrep{8}_\textbf{A}, \irrep{8}_\textbf{S}) \\ 0 & \quad (\irrep{10}_\textbf{A}, \overline{\irrep{10}}_\textbf{A}) \\ + 1 & \quad (\irrep{27}_\textbf{S}) \end{aligned} \right. .
\end{equation}
For a particle in a color representation $\irrep{R} = \irrep{3}, \irrep{6}, \irrep{8}$, the particle-antiparticle QCD potential at tree-level can be decomposed into Coulomb potentials with coupling strengths set by equation~\eqref{eq:decomposed:potentials}. The Coulomb interaction associated to a given irreducible representation $\irrep{Q}$ will affect the perturbative annihilation process for which the initial state is in the same color representation. Computing the Sommerfeld effect for a given annihilation process therefore requires decomposing the perturbative cross section according to the color representation of the particle-antiparticle initial state. Each color channel of the cross section will then be corrected independently by its own Coulomb potential order by order in the $(l,s)$ expansion. To obtain the full Sommerfeld-corrected amplitude one has to find the irreducible representations $\irrep{Q}$ contributing at each partial wave order and the weight of their relative contribution to the process.

\subsection{Decomposing perturbative cross sections}
\label{sec:decomposing:qcd:cross:section}
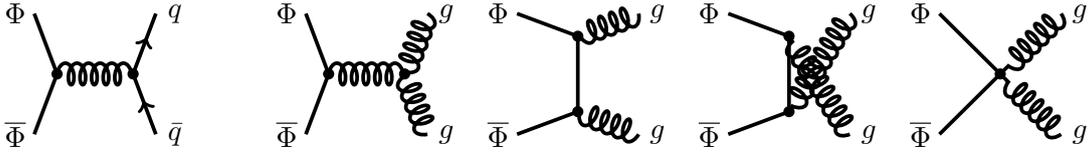
\begin{figure}
	\centering
	\begin{tikzpicture}[line width=1.4pt, scale=1]
	\draw[fermionbar] (0.8,0.8)--(0.5,0);
	\draw[fermion] (0.8,-0.8)--(0.5,0);
	\draw[fermionna] (-0.8,0.8)--(-0.5,0);
	\draw[fermionna] (-0.8,-0.8)--(-0.5,0);
	\draw[gluon] (-0.5,0)--(0.5,0);
	
	\node[vtxcircle] at (0.5,0) {};
	\node[vtxcircle] at (-0.5,0) {};
	
	\node at (-1.05,0.8) {$\Phi$};
	\node at (-1.05,-0.8) {$\overline{\Phi}$};
	\node at (1.05,0.8) {$q$};
	\node at (1.05,-0.8) {$\bar{q}$};
\end{tikzpicture}
	\hspace{6mm}
	\begin{tikzpicture}[line width=1.4pt, scale=1]
	\draw[gluon] (0.8,0.8)--(0.5,0.0);
	\draw[gluon] (0.8,-0.8)--(0.5,0.0);
	\draw[fermionna] (-0.8,0.8)--(-0.5,0);
	\draw[fermionna] (-0.8,-0.8)--(-0.5,0);
	\draw[gluon] (-0.5,0)--(0.5,0);
	
	\node[vtxcircle] at (0.5,0) {};
	\node[vtxcircle] at (-0.5,0) {};
	
	\node at (-1.05,0.8) {$\Phi$};
	\node at (-1.05,-0.8) {$\overline{\Phi}$};
	\node at (1.05,0.8) {$g$};
	\node at (1.05,-0.8) {$g$};
\end{tikzpicture}
	\hspace{-2mm}
	\begin{tikzpicture}[line width=1.4pt, scale=1]
	\draw[gluon] (0.8,0.8)--(0,0.5);
	\draw[gluon] (0.8,-0.8)--(0,-0.5);
	\draw[fermionna] (-0.8,0.8)--(0,0.5);
	\draw[fermionna] (-0.8,-0.8)--(0,-0.5);
	\draw[fermionna] (0,0.5)--(0,-0.5);
	
	\node[vtxcircle] at (0,0.5) {};
	\node[vtxcircle] at (0,-0.5) {};
	
	\node at (-1.05,0.8) {$\Phi$};
	\node at (-1.05,-0.8) {$\overline{\Phi}$};
	\node at (1.05,0.8) {$g$};
	\node at (1.05,-0.8) {$g$};
\end{tikzpicture}
	\hspace{-2mm}
	\begin{tikzpicture}[line width=1.4pt, scale=1]
	\draw[gluon] (0.8,0.8)--(0,-0.5);
	\draw[gluon] (0.8,-0.8)--(0,0.5);
	\draw[fermionna] (-0.8,0.8)--(0,0.5);
	\draw[fermionna] (-0.8,-0.8)--(0,-0.5);
	\draw[fermionna] (0,0.5)--(0,-0.5);
	
	\node[vtxcircle] at (0,0.5) {};
	\node[vtxcircle] at (0,-0.5) {};
	
	\node at (-1.05,0.8) {$\Phi$};
	\node at (-1.05,-0.8) {$\overline{\Phi}$};
	\node at (1.05,0.8) {$g$};
	\node at (1.05,-0.8) {$g$};
\end{tikzpicture}
	\hspace{-2mm}
	\begin{tikzpicture}[line width=1.4pt, scale=1]
	\draw[gluon] (0.8,0.8)--(0.0,0.0);
	\draw[gluon] (0.8,-0.8)--(0.0,0.0);
	\draw[fermionna] (-0.8,0.8)--(0,0);
	\draw[fermionna] (-0.8,-0.8)--(0,0);
	 
	\node[vtxcircle] at (0.0,0) {};
	
	\node at (-1.05,0.8) {$\Phi$};
	\node at (-1.05,-0.8) {$\overline{\Phi}$};
	\node at (1.05,0.8) {$g$};
	\node at (1.05,-0.8) {$g$};
\end{tikzpicture}
	\caption{Feynman diagrams for the annihilation of $\Phi$ into either a quark anti-quark pair or a pair of gluons. The annihilating field $\Phi$ may be scalar, fermion or vector, however, in the case of the fermion the four-point interaction is absent.}
	\label{fig:annihilation:diagrams}
\end{figure}

In this section, we consider tree-level annihilation of a particle $\Phi$ into quarks and gluons through the strong interaction
\begin{equation}
	\Phi \, \overline{\Phi} \to q_i \, \bar{q}_j \qquad \mathrm{and} \qquad \Phi \, \overline{\Phi} \to g^a \, g^b .
\end{equation}
Since no new physics couplings are involved, the nature of the diagrams contributing to the annihilation process only depends on the spin of $\Phi$. Here, we take $\Phi$ to be either a scalar, a fermion or a vector. The Feynman diagrams for the different annihilation processes are shown in figure~\ref{fig:annihilation:diagrams}. Note that the prescription in this section and the previous section for decomposing the QCD potential and cross section is also applicable to other processes. A few more exotic examples are discussed in appendix~\ref{sec:exotic:qcd:decuplet} and~\ref{sec:exotic:qcd:triplet:octet}.

First we discuss the color structure of the amplitude for the annihilation into a quark anti-quark pair. As shown in figure~\ref{fig:annihilation:diagrams}, this process occurs through a single $s$-channel gluon exchange diagram. The corresponding amplitude is therefore proportional to the generator for the $SU(3)$ representation $\irrep{R}$ of $\Phi$:
\begin{equation} \label{eq:amplitude:propto:qq}
	\mathcal{A}^a \big|^i_j \propto (T_{\irrep{R}}^a)^i_j ,
\end{equation}
where $a$ is the index of the $s$-channel gluon and the indices $i$ and $j$ run from $1$ to the dimensionality of the $\irrep{R}$ representation, $d_{\irrep{R}}$. Only the color octet configuration of the initial state, matching the representation of the exchanged gluon, will therefore contribute to the $\Phi \, \overline{\Phi} \rightarrow q_i \, \bar{q}_j$ cross section
\begin{equation} \label{eq:decomposition:color:quarks}
	\sum_\mathrm{color} \left| A_{\irrep{R} \otimes \overline{\irrep{R}}} \right|^2 = \sum_\mathrm{color} \big| [\irrep{8}] \big|^2 ,
\end{equation}
where $\sum_\mathrm{color}$ runs over all the color indices of the external particles in the amplitude.\footnote{Note that this result only holds when annihilation occurs through an $s$-channel gluon. In some non-minimal models, $\Phi$ can also annihilate into a quark anti-quark pair through the $t$-channel exchange of a new particle. We discuss this scenario in more detail in appendix~\ref{sec:exotic:qcd:triplet:triplet}.} The octet representation $[\irrep{8}]$ is antisymmetric for the decomposition of self-conjugate representations, like the $\irrep{8} \otimes \irrep{8}$, because of the CP nature of the exchanged gluon~\cite{Kauth:2011vg}.

Annihilation processes into gluons have a more complex color structure. As can be seen in figure~\ref{fig:annihilation:diagrams}, four different processes now contribute to the annihilation cross section, each with a different kinematics. The amplitudes for all of these processes, however, will be proportional to a linear combination of $T^a_{\irrep{R}} \, T^b_{\irrep{R}}$ and $T^b_{\irrep{R}} \, T^a_{\irrep{R}}$ where $a,b$ are the color indices of the final state gluons. In full generality, the amplitude can then be written as
\begin{equation} \label{eq:amplitude:propto:gg}
	\mathcal{A}^{ab} \big|^i_j = \alpha \left\{ T_{\irrep{R}}^a, T_{\irrep{R}}^b \right\}^i_j + \beta \left[ T_{\irrep{R}}^a, T_{\irrep{R}}^b \right]^i_j ,
\end{equation}
where $\alpha, \beta$ are factors that contain the kinematic dependence. As underlined in section~\ref{sec:partial:wave:expansion}, this expression drastically simplifies when the amplitude is expanded into $(l, s)$ states. For a given $(l, s)$ initial state, CP conservation enforces
\begin{equation} \label{eq:color:amplitude:symmetricity}
	\mathcal{A}^{ab} \big|^i_j = (-1)^{l+s} \mathcal{A}^{ba} \big|^i_j . 
\end{equation}
The annihilation amplitude will therefore be proportional to the anticommutator of the $T^a_{\irrep{R}}$ for  even $l + s$ and to the commutator for odd $l + s$.\footnote{Note that this result differs from the one in appendix~\ref{sec:xsec:partial:wave} of~\cite{deSimone:2014pda} and from~\cite{Ibarra:2015nca,Giacchino:2015hvk} which assume proportionality of the amplitude to the anticommutator for all values of the angular momentum.} This simplification allows us to decompose amplitudes and therefore cross sections into states of definite color independently of the kinematics of the process.

We now decompose the $\Phi \, \overline{\Phi} \rightarrow g^a \, g^b$ amplitude into contributions from initial state configurations with a definite color. As in section~\ref{sec:decomposing:qcd:potential}, we consider particle-antiparticle annihilation with $\irrep{R} = \irrep{3}, \irrep{6}, \irrep{8}$. For amplitudes proportional to $[T^a_{\irrep{R}}, T^b_{\irrep{R}}]$, we can write
\begin{equation} \label{eq:amplitude:propto:gg:odd}
	\mathcal{A}^{ab} \big|^i_j \propto [T^a_{\irrep{R}}, T^b_{\irrep{R}}]^i_j = i f^{abc} (T^c_{\irrep{R}})^i_j .
\end{equation}
As for annihilation into $q \, \bar{q}$, the amplitudes here are proportional to linear combinations of the generators of the $\irrep{R}$ representation and therefore receive contributions from color octet configurations only
\begin{equation} \label{eq:decomposition:color:gluons:odd}
	\sum_\mathrm{color} \left| A_{\irrep{R} \otimes \overline{\irrep{R}}} \right|^2 = \sum_\mathrm{color} \big| [\irrep{8}] \big|^2 .
\end{equation}
In here the $[\irrep{8}]$ is in the antisymmetric representation for the decomposition of self-conjugate representations because of the CP-odd nature of the amplitude as described in equation~\eqref{eq:color:amplitude:symmetricity}.

For terms proportional to $\left\{T^a_{\irrep{R}}, T^b_{\irrep{R}} \right\}$, the amplitude decomposition depends on $\irrep{R}$. We derive the coefficients associated to the different color representations of the initial state by decomposing the amplitude into irreducible tensors as outlined in~\cite{Georgi:1999wka}. The details of the decomposition of a given tensor for the processes and representations we are considering are presented in appendix~\ref{sec:color:decomposition:amplitude}. Applying the corresponding results to the $\left\{ T^a_{\irrep{R}}, T^b_{\irrep{R}} \right\}^i_j$ tensor allows to express the amplitude as
\begin{equation} \label{eq:amplitude:propto:gg:even}
	\mathcal{A}^{ab} \big|^i_j = \sum_{\irrep{Q}} \, [\irrep{Q}]^{ab} \big|^i_j \, ,
\end{equation}
for $\irrep{R} \otimes \overline{\irrep{R}} = \bigoplus_{\irrep{Q}} \irrep{Q}$ and where $[\irrep{Q}]^{ab}$ represents the amplitude associated to an initial state in the color representation $\irrep{Q}$. Since the contributions from the different $\irrep{Q}$ initial states are orthogonal, the squared amplitude will be of the form
\begin{equation} \label{eq:decomposition:color:gluons:irreps}
	\sum_\mathrm{color} \big| A_{\irrep{R} \otimes \overline{\irrep{R}}} \big|^2 = \sum_{\irrep{Q}} \left[\sum_\mathrm{color} \big| [\irrep{Q}] \big|^2 \right] .
\end{equation}
For the $\Phi \, \overline{\Phi} \rightarrow g^a \, g^b$ process that we consider here, using equations~\eqref{eq:decomposition:color:fundamental},~\eqref{eq:decomposition:color:symmetric} and~\eqref{eq:decomposition:color:adjoint} for terms proportional to $\left\{ T^a_{\irrep{R}}, T^b_{\irrep{R}} \right\}^i_j$, we obtain the magnitude of the contributions from the different color states to the total amplitude. For $\irrep{R} = \irrep{3}, \irrep{6}, \irrep{8}$, we have
\begin{equation} \label{eq:decomposition:color:gluons:even}
	\begin{aligned}
		\sum_\mathrm{color} \big| A_{\irrep{3} \otimes \bar{\irrep{3}}} \big|^2 & = \frac{7}{2} \sum_\mathrm{color} \big| [\irrep{1}] \big|^2 = \frac{7}{5} \sum_\mathrm{color} \big| [\irrep{8}] \big|^2 \\
		\sum_\mathrm{color} \big| A_{\irrep{6} \otimes \bar{\irrep{6}}} \big|^2 & = \frac{31}{5} \sum_\mathrm{color} \big| [\irrep{1}] \big|^2 = \frac{155}{49} \sum_\mathrm{color} \big| [\irrep{8}] \big|^2 = \frac{155}{81} \sum_\mathrm{color} \big| [\irrep{27}] \big|^2 \\
		\sum_\mathrm{color} \big| A_{\irrep{8} \otimes \irrep{8}} \big|^2 & = 6 \sum_\mathrm{color} \big| [\irrep{1}_\mathrm{\textbf{S}}] \big|^2 = 3 \sum_\mathrm{color} \big| [\irrep{8}_\mathrm{\textbf{S}}] \big|^2 = 2 \sum_\mathrm{color} \big| [\irrep{27}_\mathrm{\textbf{S}}] \big|^2. 
	\end{aligned}
\end{equation}
These results for the triplet and the octet agree with the ones obtained for the $s$-wave in~\cite{deSimone:2014pda,Liew:2016hqo}. The results for the sextet and the more exotic decompositions discussed in appendix~\ref{sec:exotic:qcd:decuplet} and~\ref{sec:exotic:qcd:triplet:octet} are novel and can also be used to extend the scope of the bound state calculations of~\cite{Liew:2016hqo} as described in~\cite{ElHedri:2017nny}. 

\subsection{Sommerfeld corrections}
\label{sec:decomposing:qcd:sommerfeld}
Combining the results from sections~\ref{sec:decomposing:qcd:potential} and~\ref{sec:decomposing:qcd:cross:section}, the Sommerfeld-corrected cross sections for the annihilation of two colored states in the representations $\irrep{R}$ and $\overline{\irrep{R}}$ can be decomposed as
\begin{equation}
	\sigma^{(S)} = \sum_{\irrep{Q}} \kappa_{\irrep{Q}} \, \sigma^{(S)}_{\mathrm{C}} \left[ \alpha_{\irrep{Q}} \right] ,
\end{equation}
where $\irrep{R} \otimes \overline{\irrep{R}} = \bigoplus_{\irrep{Q}} \irrep{Q}$. $\sigma^{(S)}_C \left[ \alpha_{\irrep{Q}} \right]$ is the Sommerfeld-corrected cross section for a Coulomb potential with coupling strength $A = \alpha_{\irrep{Q}}$, which can be computed by combining equations~\eqref{eq:total:cross:section:partial:waves} and~\eqref{eq:sommerfeld:enhanced:sqmatrix}. $\kappa_{\irrep{Q}}$ is the relative magnitude of the contribution of the $\irrep{Q}$ initial state to the annihilation amplitude, defined as
\begin{equation}
	\sum_\mathrm{color} \big| [\irrep{Q}] \big|^2 = \kappa_{\irrep{Q}} \sum_\mathrm{color} \big| A_{\irrep{R} \otimes \overline{\irrep{R}}} \big|^2 .
\end{equation}
As described in section~\ref{sec:decomposing:qcd:cross:section}, the $\kappa_{\irrep{Q}}$ weights depend not only on the color representation of the initial state, but also on its $(l, s)$ quantum numbers and on the process considered. Notably, for $\Phi \, \overline{\Phi} \rightarrow g^a \, g^b$, states with even and odd $l+s$ are respectively proportional to the anticommutator and the commutator of the color generators and therefore have different $\kappa_{\irrep{Q}}$ factors. In what follows, we will therefore consider cross sections associated to an individual $(l, s)$ particle-antiparticle initial state in the $\irrep{R} \otimes \overline{\irrep{R}}$ representation.

Reading off $\alpha_{\irrep{Q}}$ from equation~\eqref{eq:decomposed:potentials} and $\kappa_{\irrep{Q}}$ from equations~\eqref{eq:decomposition:color:quarks}, \eqref{eq:decomposition:color:gluons:odd} and \eqref{eq:decomposition:color:gluons:even}, for $\irrep{R} = \irrep{3}, \irrep{6}, \irrep{8}$, the Sommerfeld-corrected cross sections are
\begin{equation} \label{eq:colored:sommerfeld:corrections}
	\begin{aligned}
		\sigma^{(S)}_{\irrep{3} \otimes \overline{\irrep{3}} \to q \, \overline{q}} & = \sigma^{(S)}_C \left[ - \frac{\alpha_s}{6} \right] \\
		\sigma^{(S)}_{\irrep{3} \otimes \overline{\irrep{3}} \to g \, g} & = \begin{cases} \frac{2}{7} \sigma^{(S)}_C \left[ \frac{4 \alpha_s}{3} \right] + \frac{5}{7} \sigma^{(S)}_C \left[ - \frac{\alpha_s}{6} \right] & \mathrm{even} \,\, l + s \\ \sigma^{(S)}_C \left[ - \frac{\alpha_s}{6} \right] & \phantom{!} \mathrm{odd} \,\, l + s \end{cases} \\
		\sigma^{(S)}_{\irrep{6} \otimes \overline{\irrep{6}} \to q \, \overline{q}} & = \sigma^{(S)}_C \left[ \frac{11 \alpha_s}{6} \right] \\
		\sigma^{(S)}_{\irrep{6} \otimes \overline{\irrep{6}} \to g \, g} & = \begin{cases} \frac{5}{31} \sigma^{(S)}_C \left[ \frac{10 \alpha_s}{3} \right] + \frac{49}{155} \sigma^{(S)}_C \left[ \frac{11 \alpha_s}{6} \right] + \frac{81}{155} \sigma^{(S)}_C \left[ - \frac{2 \alpha_s}{3} \right] & \mathrm{even} \,\, l + s \\ \sigma^{(S)}_C \left[ \frac{11 \alpha_s}{6} \right] & \phantom{!} \mathrm{odd} \,\, l + s \end{cases} \\
		\sigma^{(S)}_{\irrep{8} \otimes \irrep{8} \to q \, \overline{q}} & = \sigma^{(S)}_C \left[ \frac{3 \alpha_s}{2} \right] \\
		\sigma^{(S)}_{\irrep{8} \otimes \irrep{8} \to g \, g} & = \begin{cases} \frac{1}{6} \sigma^{(S)}_C \left[ 3 \alpha_s \right] + \frac{1}{3} \sigma^{(S)}_C \left[ \frac{3 \alpha_s}{2} \right] + \frac{1}{2} \sigma^{(S)}_C\left[ - \alpha_s \right] & \mathrm{even} \,\, l + s \\ \sigma^{(S)}_C \left[ \frac{3 \alpha_s}{2} \right] & \phantom{!} \mathrm{odd} \,\, l + s \end{cases} .
	\end{aligned}
\end{equation}
The Coulomb cross sections $\sigma^{(S)}_C\left[ \alpha \right]$ can be readily obtained by plugging the right value for $\alpha$ into the analytic expressions in section~\ref{sec:sommerfeld:partialwaves}. The final analytic expressions for the Sommerfeld-corrected cross sections can be found by combining equations~\eqref{eq:total:cross:section:partial:waves} and~\eqref{eq:sommerfeld:enhanced:sqmatrix}.

The results in this section have been based on the assumption that annihilation always involves initial states of definite color. However, as argued in~\cite{Berger:2008ti,Baer:1998pg}, rapid interactions of the annihilating particles with the gluons in the thermal bath may prevent the initial state to be in a definite color channel. The importance of this effect is unclear since the time scale may be of the same order as the Sommerfeld effect. Its impact on the cross section can be bounded by considering an extreme scenario where annihilation always involves color-averaged initial states. As mentioned in section~\ref{sec:decomposing:qcd:cross:section}, for annihilation processes into two quarks or into two gluons with odd $l + s$, the initial state has to always be a color octet. These processes are therefore not modified by color-averaging. For annihilation into gluon pairs with even $l + s$ on the other hand, one has to use the averaged equivalent of equation~\eqref{eq:decomposed:potentials} for the QCD potential. This new potential can be straightforwardly obtained from equation~\eqref{eq:colored:sommerfeld:corrections} by averaging over the different channels. We then obtain
\begin{equation} \label{eq:decomposed:potentials:averaged}
	V_{\irrep{3} \otimes \overline{\irrep{3}} \to g \, g}^\mathrm{avg} = - \frac{11}{42} \frac{\alpha_s}{r}, \quad \qquad V_{\irrep{6} \otimes \overline{\irrep{6}} \to g \, g}^\mathrm{avg} = - \frac{143}{186} \frac{\alpha_s}{r}, \quad \qquad V_{\irrep{8} \otimes \irrep{8} \to g \, g}^\mathrm{avg} = - \frac{1}{2} \frac{\alpha_s}{r} .
\end{equation}
This leads to modified Sommerfeld-correction factors for the annihilation into two gluons with even $l+s$ as
\begin{equation} \label{eq:colored:sommerfeld:corrections:averaged}
	\begin{aligned}
		\sigma^{(S), \, \mathrm{avg}}_{\irrep{3} \otimes \overline{\irrep{3}} \to g \, g} & = \sigma^{(S)}_C \left[ \frac{11 \alpha_s}{42} \right]  \\
		\sigma^{(S), \, \mathrm{avg}}_{\irrep{6} \otimes \overline{\irrep{6}} \to g \, g} & = \sigma^{(S)}_C \left[ \frac{143 \alpha_s}{186} \right]  \\
		\sigma^{(S), \, \mathrm{avg}}_{\irrep{8} \otimes \irrep{8} \to g \, g} & = \sigma^{(S)}_C \left[ \frac{\alpha_s}{2} \right] .
	\end{aligned}
\end{equation}
In the following section, we assume that the annihilation processes occur through definite color channels. We emphasize however that the Sommerfeld-corrected cross sections in the color-averaged scenario can also be readily calculated using our formalism.

\section{Annihilation in the colored dark sector}
\label{sec:colored:dark:sector}
In the previous two sections we described how to analytically calculate Sommerfeld corrections for the annihilation of colored particles including higher order partial waves. We are now ready to apply these prescriptions to actual colored dark sectors. We imagine that the dark sector consists of a single dark matter particle which is a singlet under the Standard Model gauge groups. Furthermore the dark sector has a colored particle $\Phi$ with arbitrary spin --- scalar, fermion or vector --- and with an arbitrary representation under $SU(3)$. We then introduce a small coupling between DM and $\Phi$ ensuring chemical and thermal equilibrium between both particles. The details and phenomenology of this construction are described in an upcoming accompanying paper~\cite{ElHedri:2017nny}, here we only focus on the annihilation of the colored particle $\Phi$. We note that in these types of constructions the relic abundance is completely determined by the annihilation rate of the colored particle. 

These simple models have been introduced for illustrative purposes. We emphasize however that the methods detailed in this paper are applicable to the annihilation of colored particles in any kind of dark sector. In the rest of this section, we introduce a set of simplified models for $\Phi$ and compute the associated Sommerfeld corrections. 

\subsection{Simplified models}
\label{sec:simplified:models}
We consider scenarios where $\Phi$ is either a real or complex scalar, Dirac or Majorana fermion or a real or complex vector boson. The kinetic and mass terms for $\Phi = \left\{S, \psi, V\right\}$ in the complex scalar, Dirac fermion and complex vector models are then
\begin{equation} \label{eq:lagrangians:phi}
	\begin{aligned}
		\mathcal{L}_S & = \left[D_{\mu,ij}^{\vphantom{\mu}} S_j\right]^\dagger \left[D_{ij}^\mu S_j\right] - m_S^2 \, S_i^\dagger S_i \\
		\mathcal{L}_\psi & = \bar{\psi}_i \slashed{D}_{ij} \psi_j - m_\psi \bar{\psi}_i \psi_i \\
		\mathcal{L}_V & = -\frac{1}{2} {V_{\mu\nu,i}}^\dagger V^{\mu\nu}_i - i g_s {V_i^\mu}^\dagger (T^a_{\irrep{R}})_{ij} V_j^\nu G_{\mu\nu}^a + m_V^2 V_\mu^\dagger V^\mu ,
	\end{aligned}
\end{equation}
where $i, j$ are color indices and the $T^a_{\irrep{R}}$ matrices are the generators for the color representation $\irrep{R}$ of $\Phi$. To obtain the Lagrangians for real scalars, Majorana fermions and real vectors each of the individual terms need to be multiplied by a factor one half. The covariant derivatives and field strength are given by
\begin{equation}
	\begin{aligned}
		V^{\mu\nu}_i & = D^\mu_{ij} V^\nu_j - D^\nu_{ij} V^\mu_j \\
		D_{\mu, ij} & = \partial_\mu \delta_{ij} - i g_s G_\mu^a (T^a_{\irrep{R}})_{ij} .
	\end{aligned}
\end{equation}
Note that the Lagrangian for vectors can also include anomalous terms~\cite{Blumlein:1996qp,Hewett:1993ks} that we chose not to include in this study. The implications of using a St\"uckelberg mass term for vector $\Phi$, especially on perturbative unitarity, are discussed in our companion paper~\cite{ElHedri:2017nny}.

We list here the analytic cross sections for the pair-annihilation $\Phi$ to $q \, \bar{q}$ and $g \, g$. The total annihilation cross sections for $\Phi = S, \psi, V$ are
\begin{equation} \label{eq:analytic:xsec:annihilation}
	\begin{aligned}
		\sigma \! \left( S \, S \! \to \! q \, \bar{q} \right) = & \frac{2 \pi \alpha_s^2}{3 s} \frac{C_2(\irrep{R})}{d_{\irrep{R}}} \beta_S \\
		\sigma \! \left( S \, S \! \to \! g \, g \right) = & \frac{2 \pi \alpha_s^2}{3 s^3} \frac{C_2(\irrep{R})}{d_{\irrep{R}} \beta_S^2} \! \Bigg[ \! C_2(\irrep{G}) \left( s \beta_S (10 m_S^2 - s) - 24 m_S^4 \log \frac{1 \! + \! \beta_S}{1 \! - \! \beta_S} \right) \Bigg. \\
		& \Bigg. + 6 C_2(\irrep{R}) \left( s \beta_S (s + 4 m_S^2) + m_S^2 (8 m_S^2 - 4 s) \log \frac{1 \! + \! \beta_S}{1 \! - \! \beta_S} \right) \!\! \Bigg] \\
		\sigma \! \left( \psi \, \bar{\psi} \! \to \! q \, \bar{q} \right) = & \frac{2 \pi \alpha_s^2}{3 s} \frac{C_2(\irrep{R})}{d_{\irrep{R}}} \frac{1}{\beta_\psi} \left( 1 + \frac{2 m_\psi^2}{s} \right) \\
		\sigma \! \left( \psi \, \bar{\psi} \! \to \! g \, g \right) = & - \frac{2 \pi \alpha_s^2}{3 s^3} \frac{C_2(\irrep{R})}{d_{\irrep{R}} \beta_\psi^2} \! \Bigg[ \! C_2(\irrep{G}) \left( s \beta_\psi (s + 5 m_\psi^2) - 12 m_\psi^4 \log \frac{1 \! + \! \beta_\psi}{1 \! - \! \beta_\psi} \right) \Bigg. \\
		& \Bigg. + 3 C_2(\irrep{R}) \left( s \beta_\psi (s + 4 m_\psi^2) + (8 m_\psi^4 - 4 m_\psi^2 s - s^2) \log \frac{1 \! + \! \beta_\psi}{1 \! - \! \beta_\psi} \right) \!\! \Bigg] \\
		\sigma \! \left( V \, V \! \to \! q \, \bar{q} \right) = & \frac{\pi \alpha_s^2}{54 s} \frac{C_2(\irrep{R})}{d_{\irrep{R}}} \beta_V \frac{12 m_V^4 + 20 m_V^2 s + s^2}{m_V^4} \\
		\sigma \! \left( V \, V \! \to \! g \, g \right) = & \frac{2 \pi \alpha_s^2}{9 m_V^2 s^3} \frac{C_2(\irrep{R})}{d_{\irrep{R}} \beta_V^2} \! \Bigg[ \! C_2(\irrep{G}) m_V^2 \!\! \left( \! s \beta_V \! \left( 10 m_V^2 \! + \! 7 s \right) \! - \! 8 \! \left( 3 m_V^4 \! + \! s^2 \right) \log \frac{1 \! + \! \beta_V}{1 \! - \! \beta_V} \! \right) \! \Bigg. \\
		& \Bigg. \! + \! 2 C_2(\irrep{R}) \! \left( \! s \beta_V \! \left( 12 m_V^4 \! + \! 3 m_V^2 s \! + \! 4 s^2 \right) \! + \! 12 \left( 2 m_V^6 \! - \! m_V^4 s \right) \log \frac{1 \! + \! \beta_V}{1 \! - \! \beta_V} \! \right) \!\! \Bigg] .
	\end{aligned}
\end{equation}
In these expressions, the phase space factor is defined by $\beta_\Phi = \sqrt{1 - \frac{4 m_\Phi^2}{s}}$ and $C_2(\irrep{G}) = N$ is the quadratic Casimir of $SU(N)$. The annihilation cross sections are the same for real scalars, Majorana fermions and real vectors. Note that, since we directly introduced a squared mass term for $\Phi$ in the Lagrangian, the $V \, V \rightarrow q \, \bar{q}$ cross section grows as $\mathcal{O}(s)$ at large center-of-mass energies. This non-physical behavior can be corrected by introducing a Higgs-type particle. We discuss the associated effects on the phenomenology in our companion paper~\cite{ElHedri:2017nny}.

\subsection{Sommerfeld-corrected annihilation}
\label{sec:sommerfeld:corrected:annihilation}
This section shows the Sommerfeld corrections to the annihilation of colored particles for the non-relativistic velocities typical to most thermal dark matter models. Before freeze-out, dark matter and the particles it is in thermal equilibrium with are forming a thermal bath of relatively low temperatures compared to their masses. Around freeze-out, when the rate of the annihilation processes determines the dark matter relic density, the fraction $x = m / T$ is usually around $25$. This leads to typical velocities around $0.2$ using the Maxwell-Boltzmann distribution. Since, the contributions from larger velocities are exponentially suppressed, we study the effects of the Sommerfeld corrections in the thermally relevant range $0 < v < 0.5$.

We have implemented the procedure detailed in sections~\ref{sec:sommerfeld:partialwaves} and \ref{sec:sommerfeld:qcd} as well as the perturbative amplitudes for the models described in section~\ref{sec:colored:dark:sector} in a \texttt{Mathematica} notebook that is attached to this paper~\cite{ElHedri:2016pac}. This notebook also provides an interface to \texttt{micrOMEGAs}~\cite{Belanger:2014vza,Barducci:2016pcb} for the calculation of the Sommerfeld-corrected relic abundance in these models. Furthermore, note that this notebook can also be readily used to compute the Sommerfeld effect on amplitudes that are not studied here. The conventions and definitions used to compute the perturbative amplitudes are detailed in appendix~\ref{sec:xsec:partial:wave}.

\begin{figure}[!ht]
	\centering
	\includegraphics[width=0.495\textwidth]{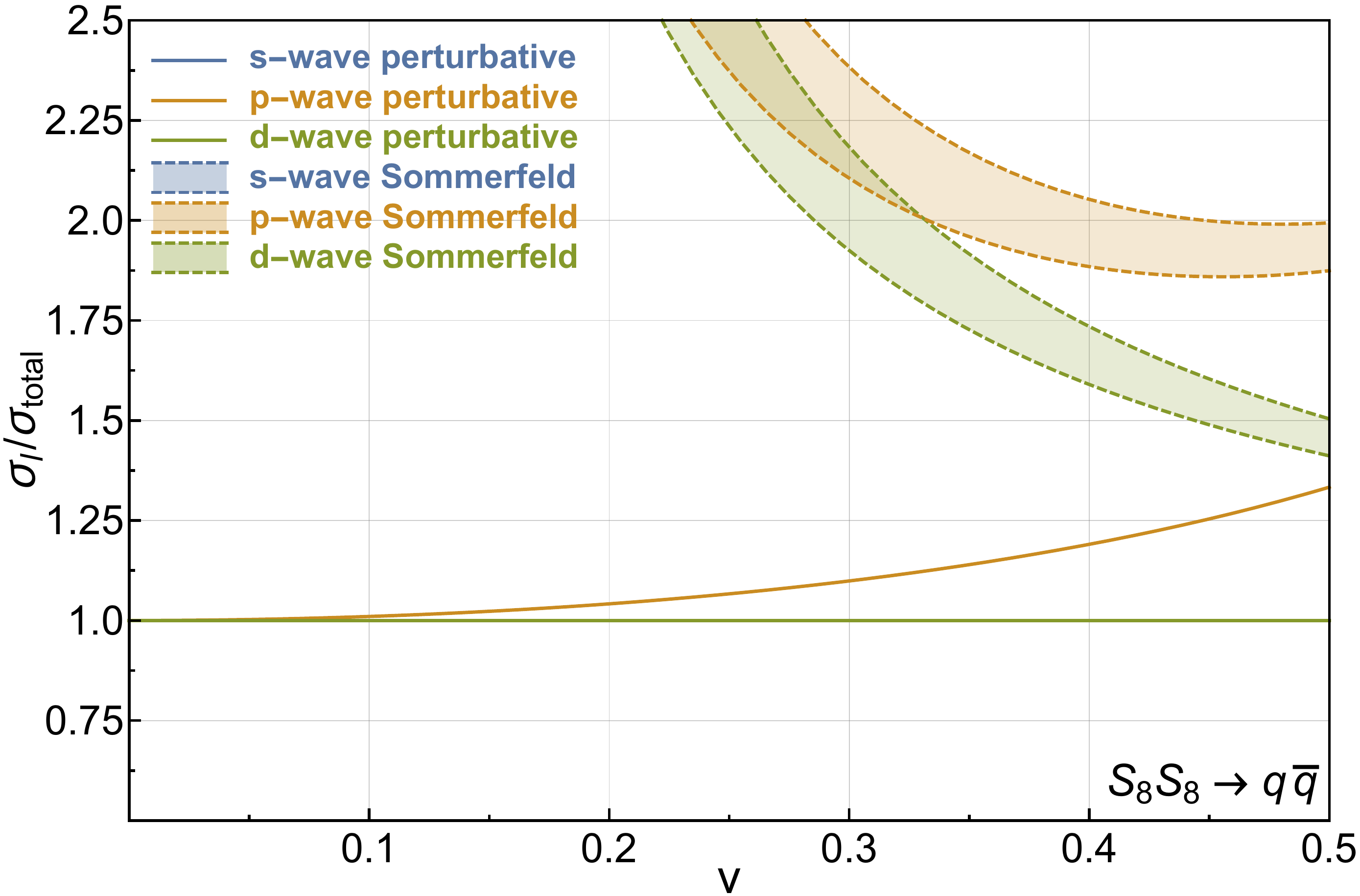}
	\includegraphics[width=0.495\textwidth]{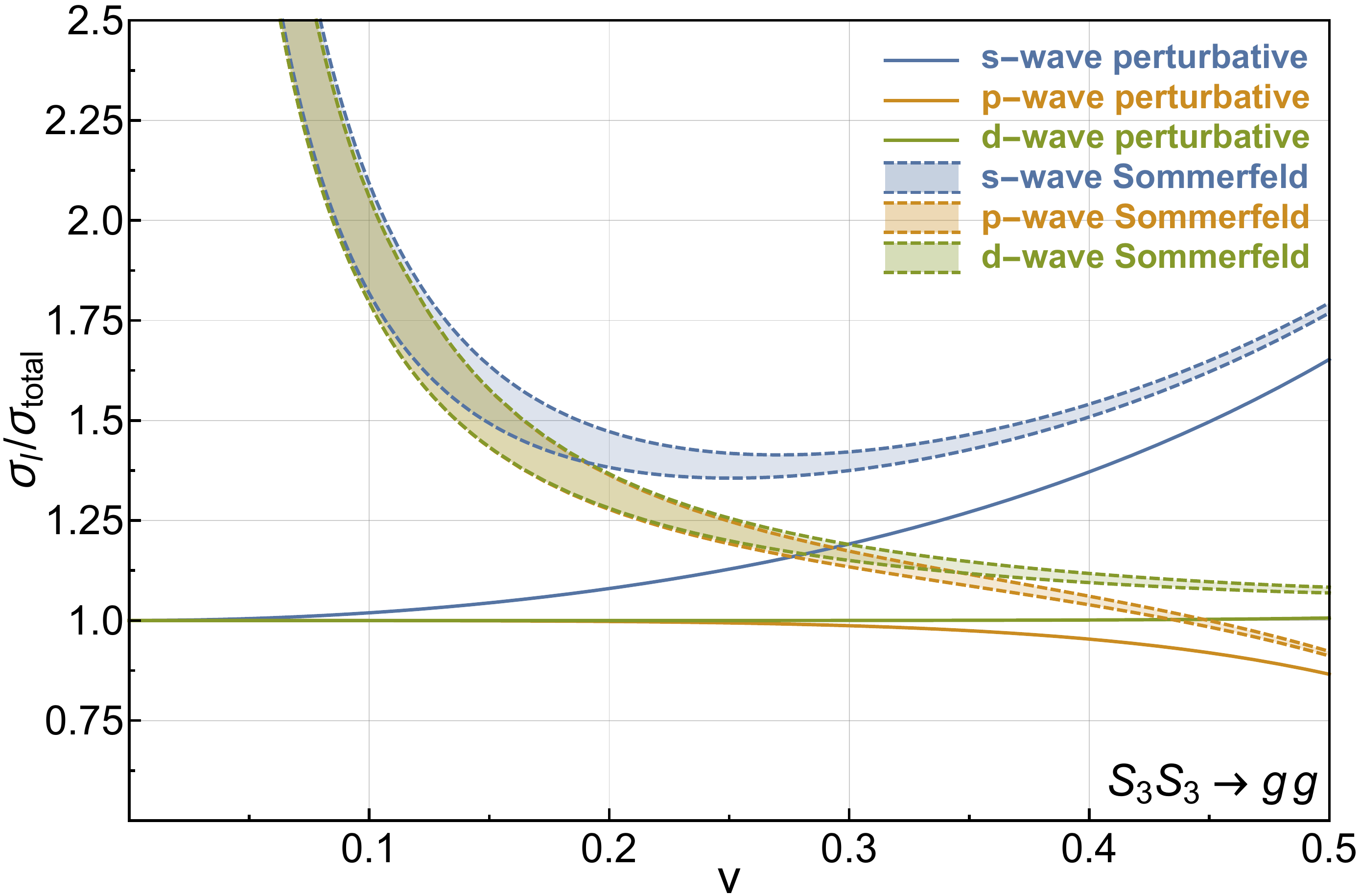}
	\includegraphics[width=0.495\textwidth]{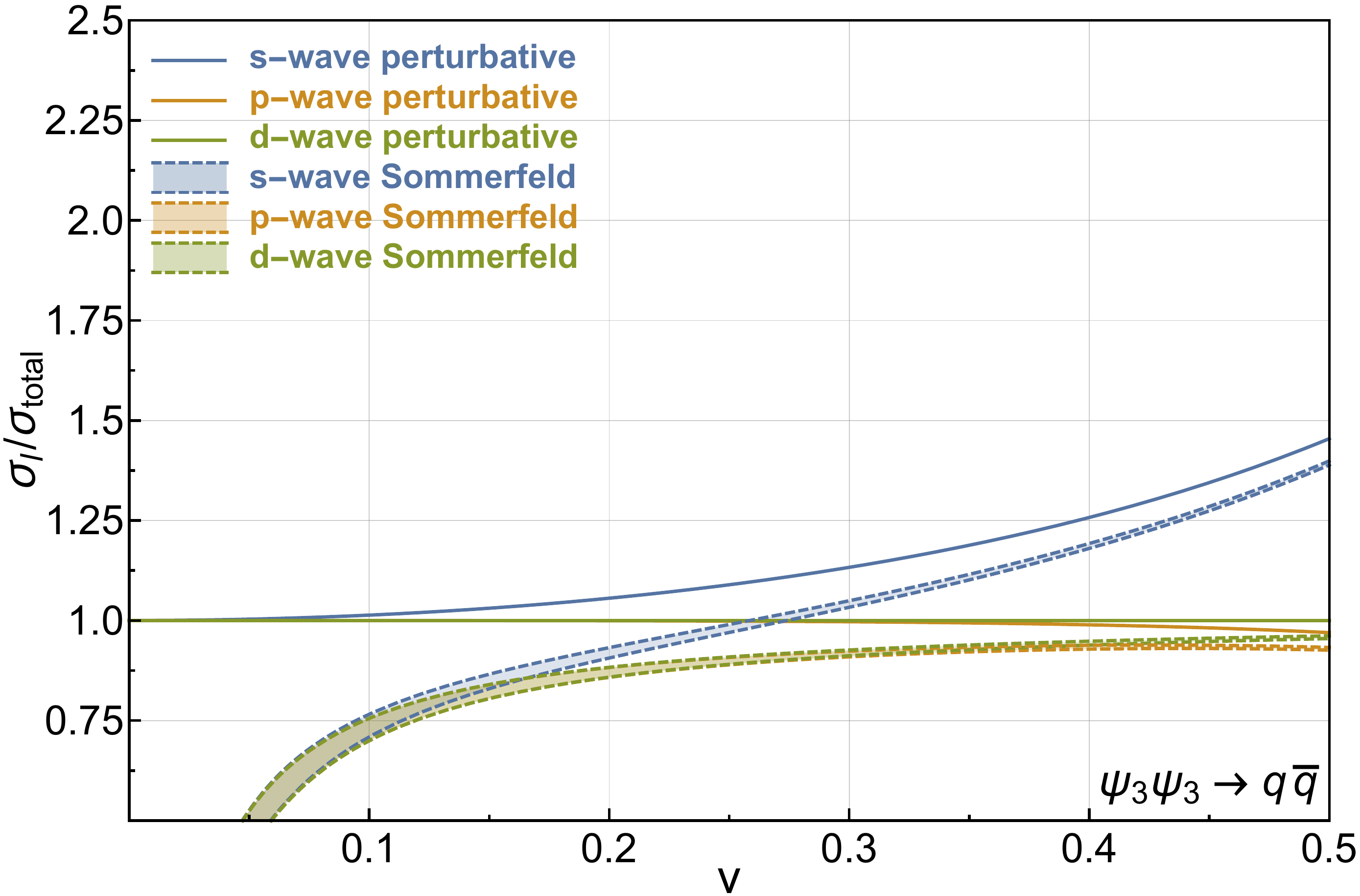}
	\includegraphics[width=0.495\textwidth]{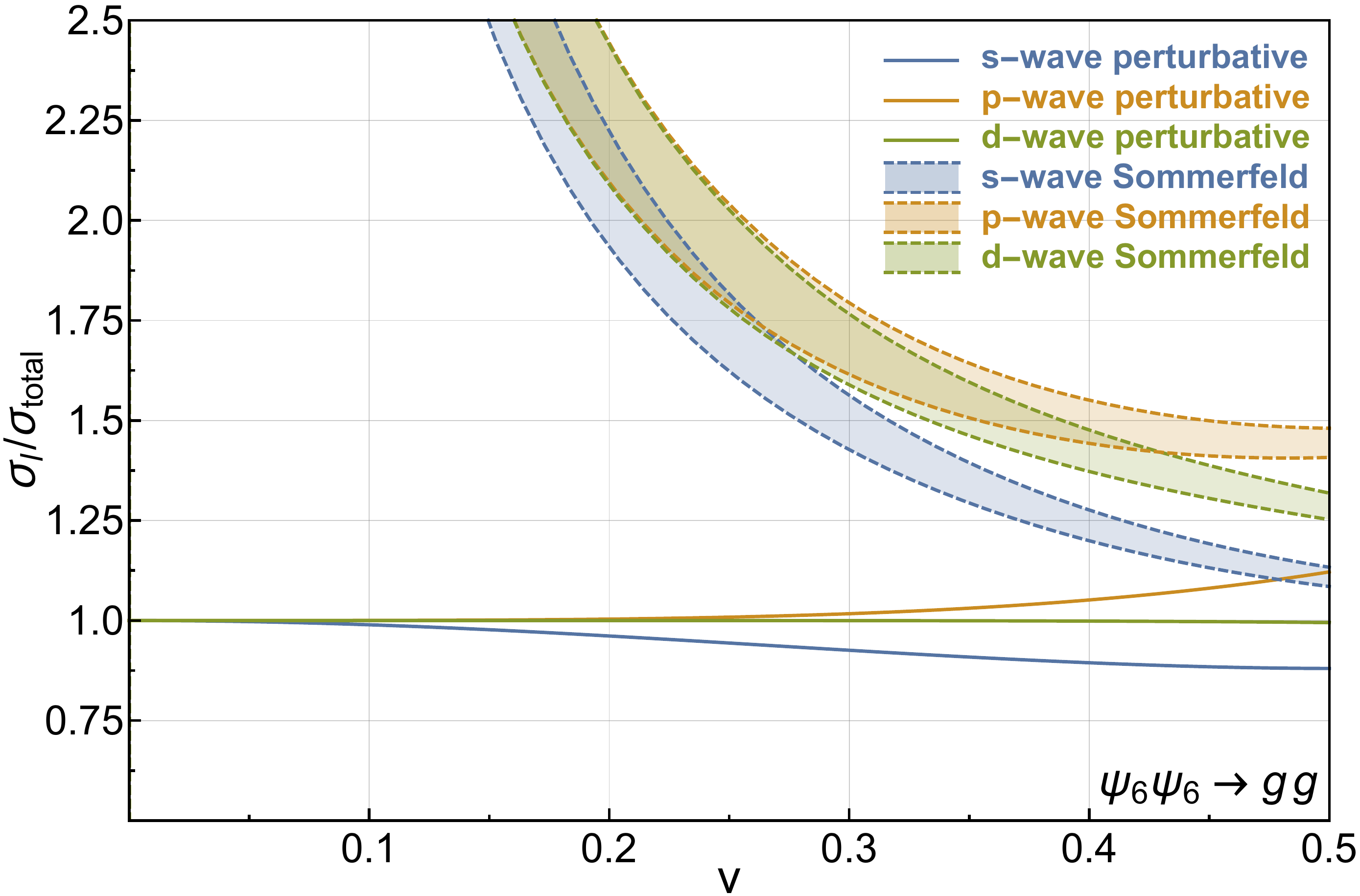}
	\includegraphics[width=0.495\textwidth]{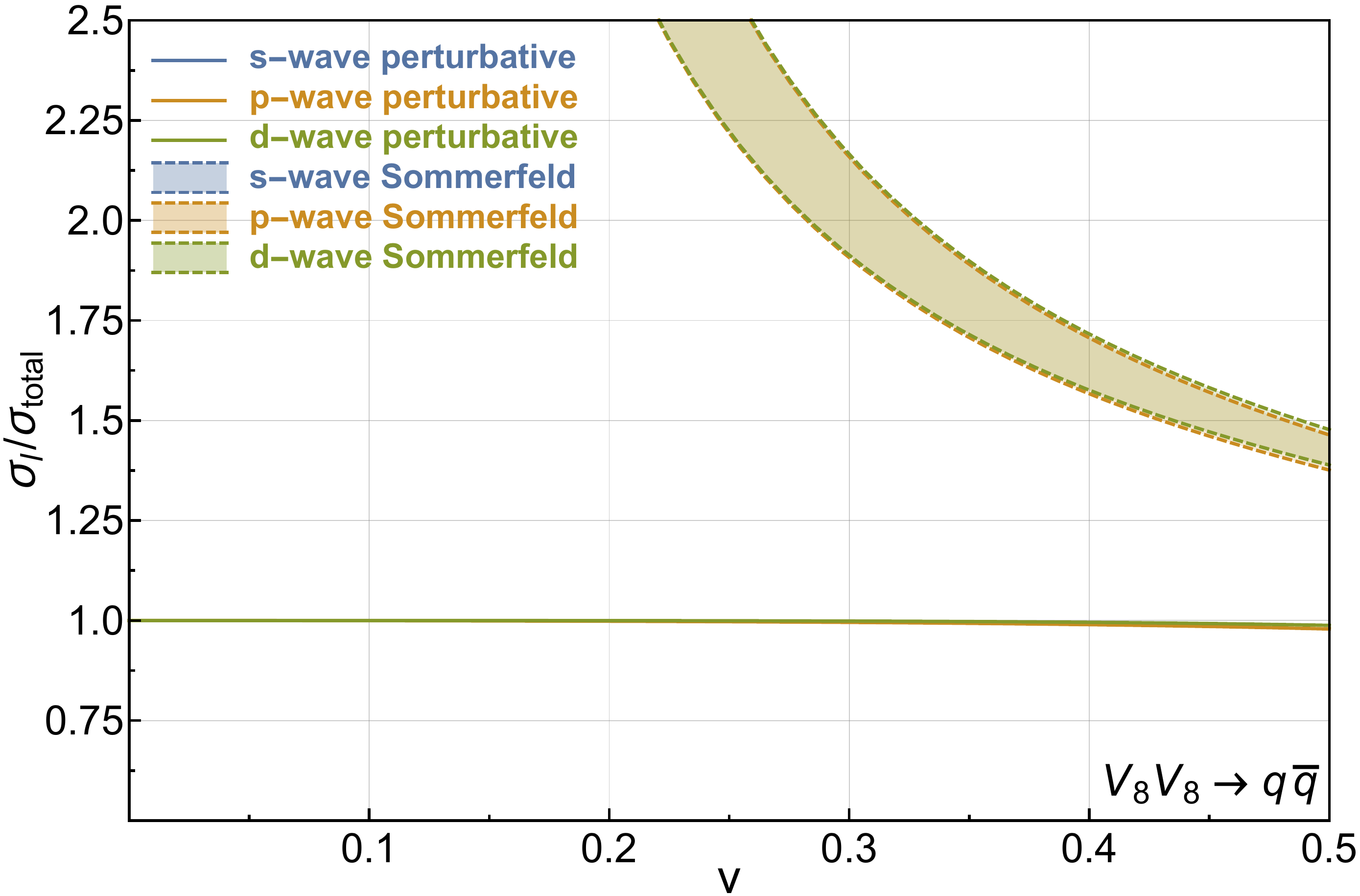}
	\includegraphics[width=0.495\textwidth]{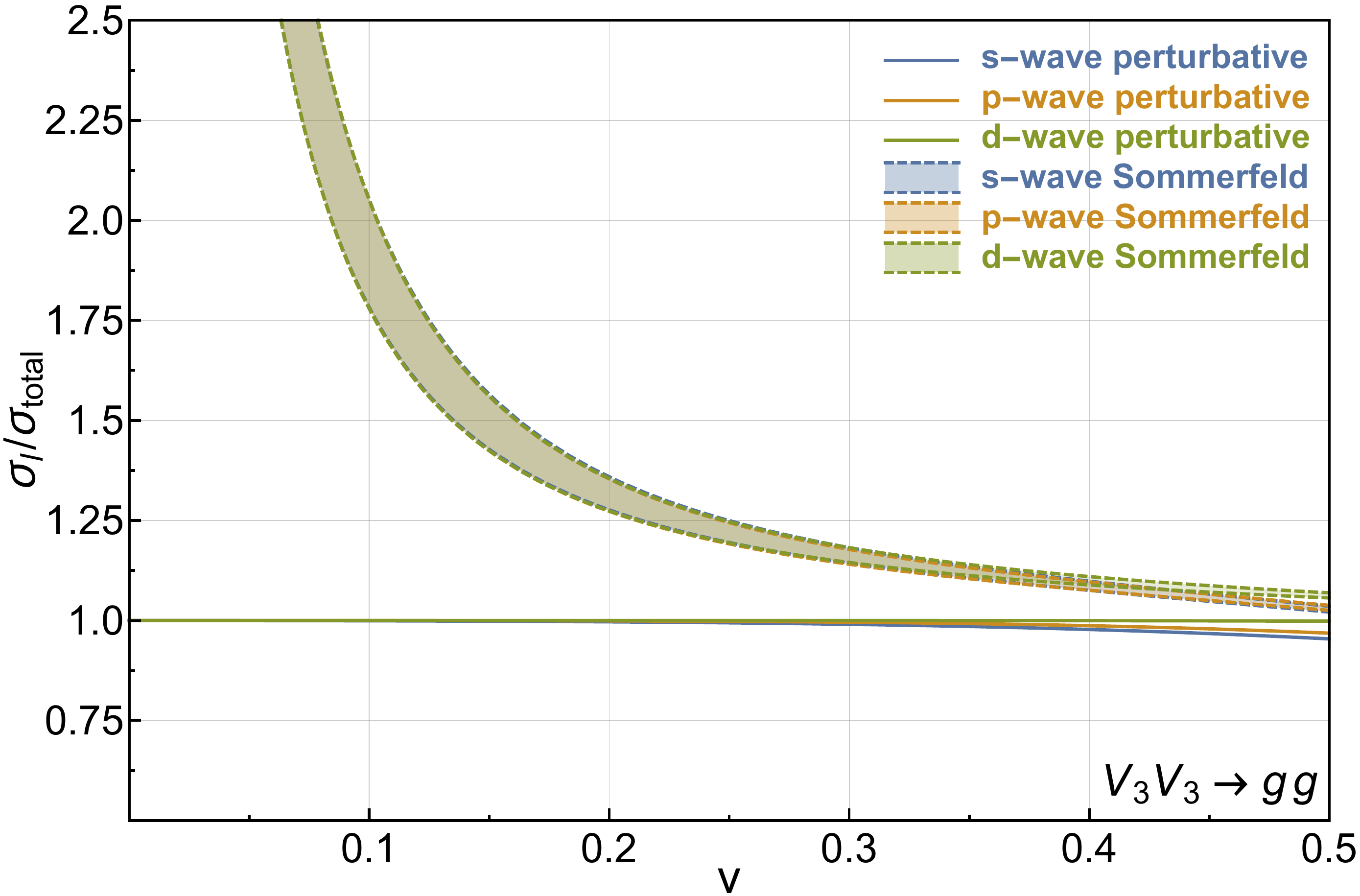}
	\caption{Ratios of the perturbative (solid lines) and Sommerfeld-corrected cross sections (dashed lines) expanded up to the $s$-wave (blue), $p$-wave (orange) and $d$-wave (green) over the exact value of the perturbative cross section. Due to the mass dependence of $\alpha_s^\mathrm{Sommerfeld} (\hat{\mu})$, the Sommerfeld-corrected ratios are shown as a band corresponding to $500~\mathrm{GeV} \leq m_\Phi \leq 2500~\mathrm{GeV}$. For each of the processes we show the results for a specific color representation, which is denoted by the subscript on the $\Phi$ fields.}
	\label{fig:annihilation:cross:section}
\end{figure}

In what follows, we consider the ratios of the partial wave expansions of the perturbative and Sommerfeld-corrected cross sections up to the $d$-wave over the exact value of the perturbative cross section. For the perturbative cross sections, we evaluate the strong coupling $\alpha_s (\mu)$ at the scale set by the mass of the annihilating particles. However when taking the ratio of the cross sections this mass dependence factors out. When computing the Sommerfeld corrections the coupling $\alpha_s^\mathrm{Sommerfeld} (\hat{\mu})$ must be evaluated at a much lower scale. This is in accordance with the scale of the soft gluons that are being exchanged. The scale is of the order of the momenta of the incoming particles that are annihilating and thus depends on the mass of the annihilating particles and their velocities. Since the scale dependence of $\alpha_s$ is significant for our range of velocities we use the precise results for the running of the strong coupling obtained in~\cite{Olive:2016xmw,Prosperi:2006hx}. 

The results for different annihilation processes are shown in figure~\ref{fig:annihilation:cross:section}. To outline the mass dependence of the Sommerfeld-corrected ratios discussed before, we plot these ratios as a band for $500~\mathrm{GeV} \leq m_\Phi \leq 2500~\mathrm{GeV}$. We first notice that, as mentioned in section~\ref{sec:sommerfeld:partialwaves}, in spite of the $\mathcal{O}(v^{-1})$ terms present at large $l$ due to Sommerfeld corrections both the perturbative and Sommerfeld-corrected cross sections converge at similar speeds with $l$. In particular, for all processes, the $d$-wave perturbative cross section is indistinguishable from the exact value up to $v \sim 0.5$. Although for colored vectors $l > 0$ contributions to the cross sections are negligible, for colored fermions and scalars, including higher order contributions leads to sizable modifications of the total cross section for both the perturbative and the Sommerfeld-corrected case. Notably, for velocities around $0.2$, which is typical for many thermal dark matter models, adding the $p$-wave contribution can lead to modifications of $\mathcal{O}(10\%)$ of the Sommerfeld-corrected cross section. Although in several models these effects can be mitigated by a cancellation between the $q \, \bar{q}$ and $g\,g$ contributions, our results highlight the importance of a rigorous computation of Sommerfeld corrections for more than one partial wave at a time.

As shown in figure~\ref{fig:annihilation:cross:section}, the Sommerfeld corrections can enhance the annihilation cross section of colored particles by up to a factor of two for typical dark matter velocities. This enhancement plays a crucial role in the phenomenology of models with a colored dark sector. In an accompanying paper~\cite{ElHedri:2017nny}, we show how relic density and collider constraints allow to derive model-independent bounds for scenarios where dark matter coannihilates with a colored dark partner.

\section{\texorpdfstring{$SU(N)$}{SU(N)} dark sectors}
\label{sec:sun:dark:sector}
Beyond the minimal models of dark matter explored in section~\ref{sec:colored:dark:sector}, extensions of the Standard Model involving exotic non-Abelian gauge groups have been strongly motivated in many BSM theories. In particular, wide classes of models such as neutral naturalness~\cite{Gherghetta:2016bcc}, hidden valleys~\cite{Juknevich:2009ji,Juknevich:2009gg}, dark radiation~\cite{Buen-Abad:2015ova,Ko:2016fcd} and glueball dark matter~\cite{Soni:2017nlm,Feng:2011ik,Boddy:2014qxa,Boddy:2014yra,Harigaya:2016nlg} often involve dark sector particles charged under a new $SU(N)$ gauge group. When this $SU(N)$ group is unbroken, dark gluon exchange between the dark sector particles leads to a long-range interaction through the same mechanism as the one described in sections~\ref{sec:sommerfeld:partialwaves} and~\ref{sec:sommerfeld:qcd} for colored particles. For sizable values of the dark $\alpha_N$ gauge coupling, this long-range interaction leads to significant Sommerfeld corrections that can be analytically approximated as in the QCD scenario. Computing the Sommerfeld effect is especially crucial when considering classes of models where $SU(N)$ is confining in the present universe~\cite{Soni:2017nlm,Feng:2011ik,Boddy:2014qxa,Boddy:2014yra,Harigaya:2016nlg}. Since in these models particles charged under $SU(N)$ are responsible for dark matter depletion, the Sommerfeld corrections are expected to significantly change the dark matter relic abundance.

In this section, we extend the methodology outlined in sections~\ref{sec:sommerfeld:partialwaves} and~\ref{sec:sommerfeld:qcd} for QCD to general $SU(N)$ dark sectors. We put special emphasis on the annihilation of messenger particles charged under both the SM and a dark gauge group, encountered in large categories of models. We discuss how to combine the Sommerfeld corrections from both potentials in these scenarios. To illustrate the relevance of our approach, we compute the Sommerfeld corrections for the model studied in~\cite{Soni:2017nlm} that involves dark fermions charged under both $SU(3)$ and $SU(N)$. 

\subsection{Color decomposition}
\label{sec:sun:color:decomposition}
In this section, we generalize the results derived in section~\ref{sec:sommerfeld:qcd} to particles charged under a new dark gauge group $SU(N)$, either in the fundamental $\irrep{F}$ or in the adjoint $\irrep{A}$ representation. As before, we consider the self-annihilation of a particle $\Phi$ into two fermions in the fundamental representation of $SU(N)$ or into two dark gauge bosons in the adjoint representation of $SU(N)$
\begin{equation}
	\Phi \, \overline{\Phi} \to Q_i \, \bar{Q}_j \qquad \mathrm{and} \qquad \Phi \, \overline{\Phi} \to G^a \, G^b .
\end{equation}
Here, we consider both $Q$ and $G$ to be massless. The procedure for computing the Sommerfeld corrections for this annihilation process is the same as the one described for $SU(3)$ in section~\ref{sec:sommerfeld:qcd}.  In particular, the leading order term of the $SU(N)$ potential can be described by a Coulomb potential obeying equation~\eqref{eq:potential:color:decomposition} and the symmetry constraints on the different partial waves arising from CP conservation are independent on $N$. 

A generalized version of equation~\eqref{eq:representation:product} can be derived by decomposing the $\irrep{F} \otimes \overline{\irrep{F}} $ and the $\irrep{A} \otimes \irrep{A}$ products to obtain the following possible representations for the $\Phi \, \overline{\Phi}$ initial state
\begin{equation} \label{eq:sun:product:decomposition}
	\begin{aligned}
		\irrep{F} \otimes \overline{\irrep{F}} & = \irrep{1} \oplus \irrep{A} \\
		\irrep{A} \otimes \irrep{A} & = \irrep{1}_\textbf{S} \oplus \irrep{A}_\textbf{A} \oplus \irrep{A}_\textbf{S} \oplus \irrep{B}_\textbf{S} \oplus \irrep{C}_\textbf{A} \oplus \overline{\irrep{C}}_\textbf{A} \oplus \irrep{D}_\textbf{S} \, .
	\end{aligned}
\end{equation}
A notable difference from the $SU(3)$ case here is the appearance of the $\irrep{B}_\textbf{S}$ representation for $N\geq 4$. The representations in equation~\eqref{eq:sun:product:decomposition} are associated with the following Young tableaux 
\begin{equation} \ytableausetup{mathmode, boxsize=1em}
	\irrep{F} = \raisebox{-0.7mm}{\begin{ytableau} {} \end{ytableau}} \qquad \irrep{A} = \raisebox{5mm}{\begin{ytableau} {} & {} \\ {} \\ {} \\ \none[\raisebox{-0.6mm}{\large \vdots}] \\ {} \end{ytableau}} \qquad \irrep{B} = \raisebox{5mm}{\begin{ytableau} {} & {} \\ {} & {} \\ \none[\raisebox{-0.6mm}{\large \vdots}] \\ {} \end{ytableau}} \qquad \irrep{C} = \raisebox{5mm}{\begin{ytableau} {} & {} & {} \\ {} \\ \none[\raisebox{-0.6mm}{\large \vdots}] \\ {} \end{ytableau}} \qquad \irrep{D} = \raisebox{5mm}{\begin{ytableau} {} & {} & {} & {} \\ {} & {} \\ {} & {} \\ \none[\raisebox{-0.6mm}{\large \vdots}] & \none[\raisebox{-0.6mm}{\large \vdots}] \\ {} & {} \end{ytableau}} \; ,
\end{equation}
where $\irrep{A}$ and $\irrep{D}$ have $N-1$ vertical boxes and $\irrep{B}$ and $\irrep{C}$ have one box less. These Young tableaux highlight the symmetry properties of the tensors belonging to the different representations and can therefore be used as guiding tools to decompose a given amplitude into contributions from different $SU(N)$ initial states. The dimensionality of all the representations as well as the quadratic Casimir and Dynkin indices~\cite{Jeon:2004rk,Okubo:1977mx} are summarized in the following table.
\begin{equation} \label{eq:sun:casimir:invariants}
	\setlength{\tabcolsep}{0.28em}
	\begin{tabular}{c | c c c c c c}
		\irrep{R} & $\irrep{1}$ & $\irrep{F}$ & $\irrep{A}$ & $\irrep{B}$ & $\irrep{C}$ & $\irrep{D}$ \\
		\hline
		$\mathrm{dim}(\irrep{R})$ & $1$ & $N$ & $N^2 \! - \! 1$ & $\frac{1}{4} (N^4 \! - \! 2 N^3 \! - \! 3 N^2)$ & $\frac{1}{4} (N^4 \! - \! 5 N^2 \! + \! 4)$ & $\frac{1}{4} (N^4 \! + \! 2 N^3 \! - \! 3 N^2)$ \\
		$C(\irrep{R})$ & $0$ & $\frac{1}{2}$ & $N$ & $\frac{1}{2} N^2 (N \! - \! 3)$ & $\frac{1}{2} N (N^2 \! - \! 4)$ & $\frac{1}{2} N^2 (N \! + \! 3)$ \\
		$C_2(\irrep{R})$ & $0$ & $\frac{N^2 - 1}{2N}$ & $N$ & $2 (N \! - \! 1)$ & $2 N$ & $2 (N \! + \! 1)$ \\
	\end{tabular}
\end{equation}

Calling the new gauge coupling $\alpha_N$, we can now use equations~\eqref{eq:potential:color:decomposition} and \eqref{eq:sun:product:decomposition} as well as the table in equation~\eqref{eq:sun:casimir:invariants} to derive the $SU(N)$ Coulomb potential associated with the different $\Phi \, \overline{\Phi}$ representations
\begin{equation} \label{eq:decomposed:potentials:sun}
	V_{\irrep{F} \otimes \overline{\irrep{F}}} = \! \frac{\alpha_N}{r} \! \left\{ \!\! \begin{aligned} - \frac{N^2 - 1}{2N} & \quad (\irrep{1}) \\ \frac{1}{2N} & \quad (\irrep{A}) \end{aligned} \right. , \quad V_{\irrep{A} \otimes \irrep{A}} = \! \frac{\alpha_N}{r} \! \left\{ \!\! \begin{aligned} - N & \quad (\irrep{1}_\textbf{S}) \\ - \frac{N}{2} & \quad (\irrep{A}_\textbf{A}, \irrep{A}_\textbf{S}) \\ - 1 & \quad (\irrep{B}_\textbf{S}) \\ 0 & \quad (\irrep{C}_\textbf{A}, \overline{\irrep{C}}_\textbf{A}) \\ 1 & \quad (\irrep{D}_\textbf{S}) \end{aligned} \right. .
\end{equation}
For the case of $N = 3$, this potential reduces to equation~\eqref{eq:decomposed:potentials}. For large $N$ the attractive terms increase, whereas the repulsive ones decrease or remain constant.

The Clebsch-Gordan coefficients for the decomposition of the annihilation cross sections can now be computed by following exactly the same steps as in section~\ref{sec:decomposing:qcd:cross:section}. The details of this calculation for the different annihilation processes as well as for even and odd $l + s$ are given in appendix~\ref{sec:color:decomposition:amplitude}. As in the $SU(3)$ case, since the $\Phi \, \overline{\Phi} \to Q_i \, \bar{Q}_j$ annihilation is mediated by an $s$-channel adjoint gauge boson, only initial states in the adjoint representation contribute to the total cross section. For the $\Phi \, \overline{\Phi} \to G^a \, G^b$ annihilation process, the CP conservation arguments described in section~\ref{sec:decomposing:qcd:cross:section} still apply and, as in equation~\eqref{eq:decomposition:color:gluons:odd}, the squared amplitude for odd $l + s$ can be written as 
\begin{equation} \label{eq:decomposition:sun:quarks:odd}
	\sum_\mathrm{color} \left| A_{\irrep{R} \otimes \overline{\irrep{R}}} \right|^2 = \sum_\mathrm{color} \big| [\irrep{A}] \big|^2 ,
\end{equation}
for all $\irrep{R}$. Similarly, for even $l + s$, the decompositions given in equations~\eqref{eq:decomposition:color:gluons:even} for the products of two fundamentals and two adjoints can be generalized to
\begin{equation} \label{eq:decomposition:sun:fundamentals:even}
	\sum_\mathrm{color} \big| A_{\irrep{F} \otimes \overline{\irrep{F}}} \big|^2 = \frac{N^2 - 2}{2} \sum_\mathrm{color} \big| [\irrep{1}] \big|^2 = \frac{N^2 - 2}{N^2 - 4} \sum_\mathrm{color} \big| [\irrep{A}] \big|^2 ,
\end{equation}
and 
\begin{equation} \label{eq:decomposition:sun:adjoints:even}
	\begin{aligned}
		\sum_\mathrm{color} \big| A_{\irrep{A} \otimes \irrep{A}} \big|^2 & = \frac{3}{4} (N^2 - 1) \sum_\mathrm{color} \big| [\irrep{1}_\textbf{S}] \big|^2 \\
		\sum_\mathrm{color} \big| A_{\irrep{A} \otimes \irrep{A}} \big|^2 & = 3 \sum_\mathrm{color} \big| [\irrep{A}_\textbf{S}] \big|^2 \\
		\sum_\mathrm{color} \big| A_{\irrep{A} \otimes \irrep{A}} \big|^2 & = \frac{3(N - 1)}{N-3} \sum_\mathrm{color} \big| [\irrep{B}_\textbf{S}] \big|^2 \\
		\sum_\mathrm{color} \big| A_{\irrep{A} \otimes \irrep{A}} \big|^2 & = \frac{3(N + 1)}{N+3} \sum_\mathrm{color} \big| [\irrep{D}_\textbf{S}] \big|^2 .
	\end{aligned}
\end{equation}
Note that these results only apply for $N \geq 4$. For $N = 3$ the contribution from the $\irrep{B}_\textbf{S}$ representation goes to zero. In the case of $N = 2$, if $\Phi$ is in the fundamental representation the $\Phi \, \overline{\Phi} \to G^a \, G^b$ process occurs only when $\Phi \, \overline{\Phi}$ is an $SU(2)$ singlet. When $\Phi$ is in the adjoint representation, only the $\Phi \, \overline{\Phi}$ states in the $\irrep{1}_\textbf{S}$ and the $\irrep{D}_\textbf{S} = \irrep{5}_\textbf{S}$ representation will contribute to the $\Phi \, \overline{\Phi} \to G^a \, G^b$ annihilation cross section. In the large-$N$ limit, on the other hand, we observe that annihilation to dark gauge bosons occurs dominantly through the adjoint channel for the annihilation of two fundamentals and splits evenly into the $\irrep{A}_\textbf{S}$, $\irrep{B}_\textbf{S}$ and $\irrep{D}_\textbf{S}$ channels for initial state particles in the adjoint representation.

\subsection{Sommerfeld corrections}
\label{sec:sun:sommerfeld:corrections}
\begin{figure}[!ht]
	\centering
	\includegraphics[width=0.495\textwidth]{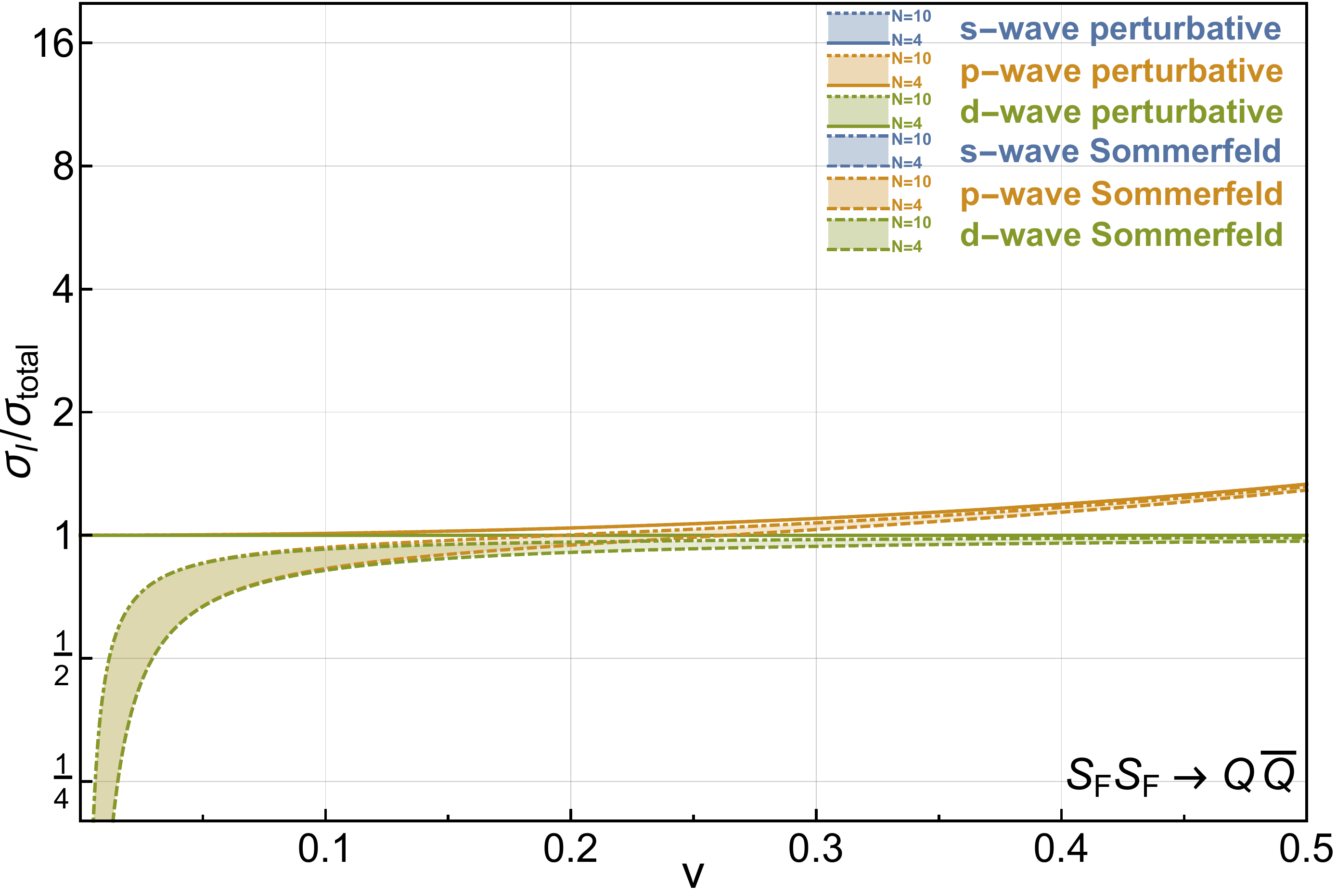}
	\includegraphics[width=0.495\textwidth]{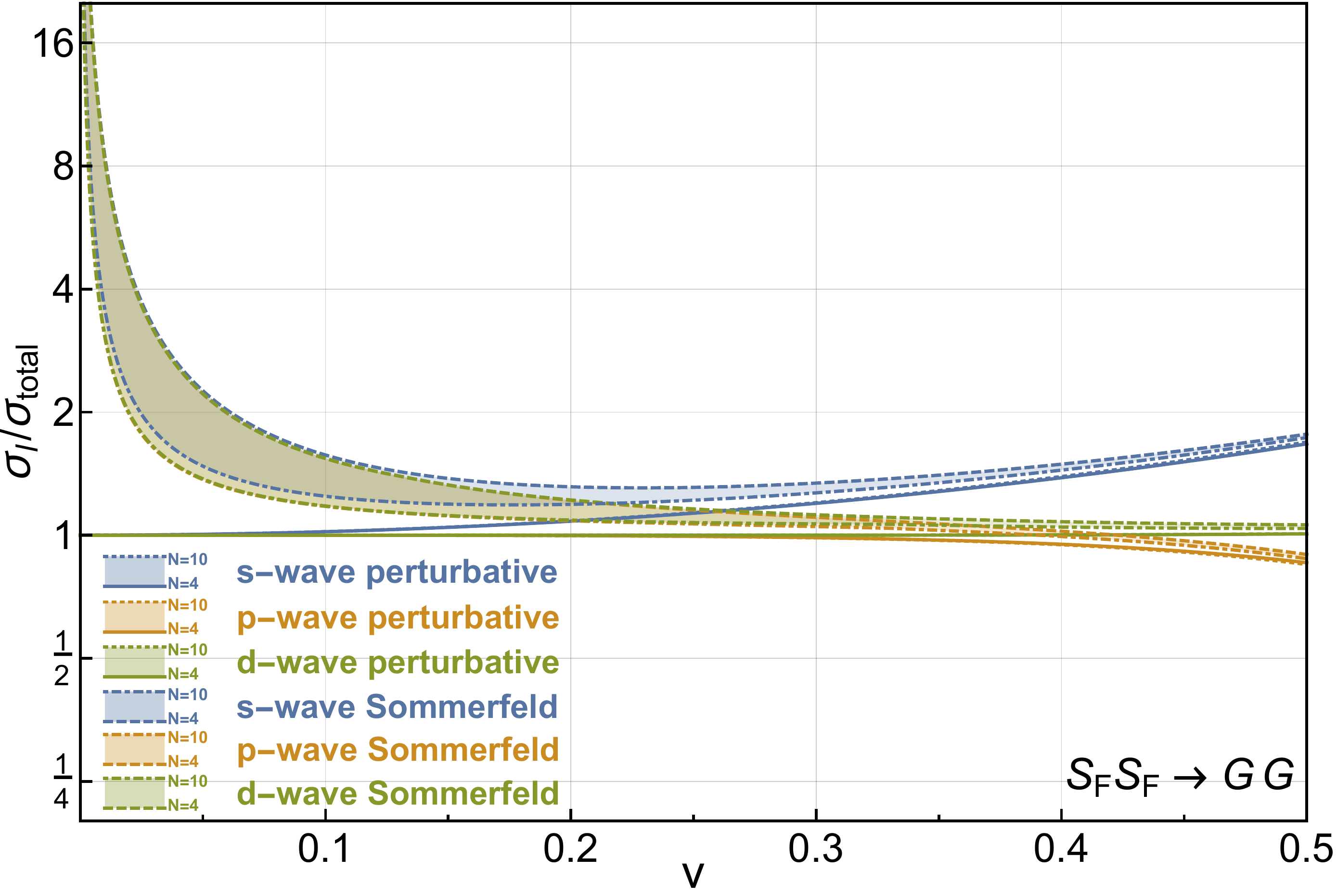}
	\includegraphics[width=0.495\textwidth]{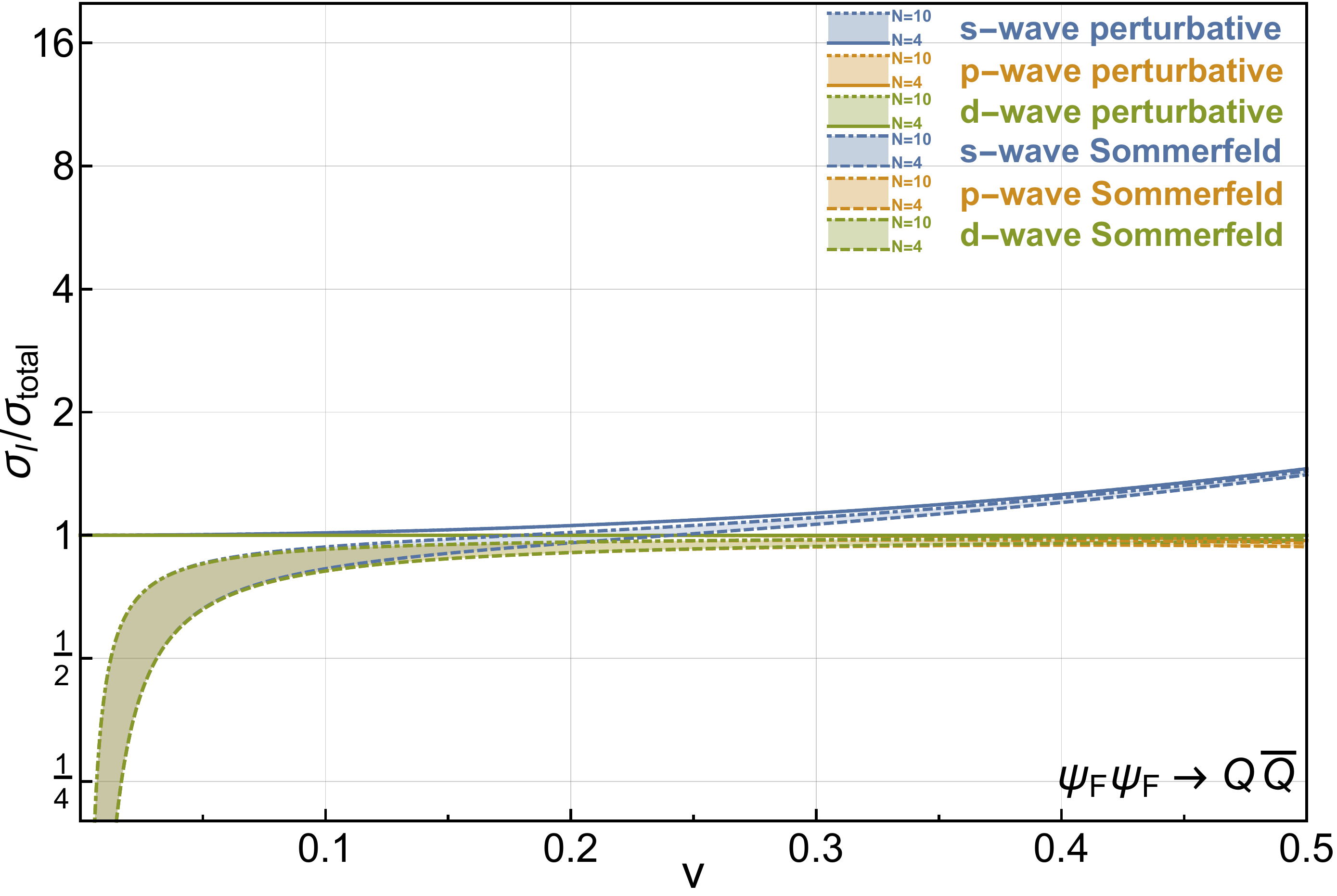}
	\includegraphics[width=0.495\textwidth]{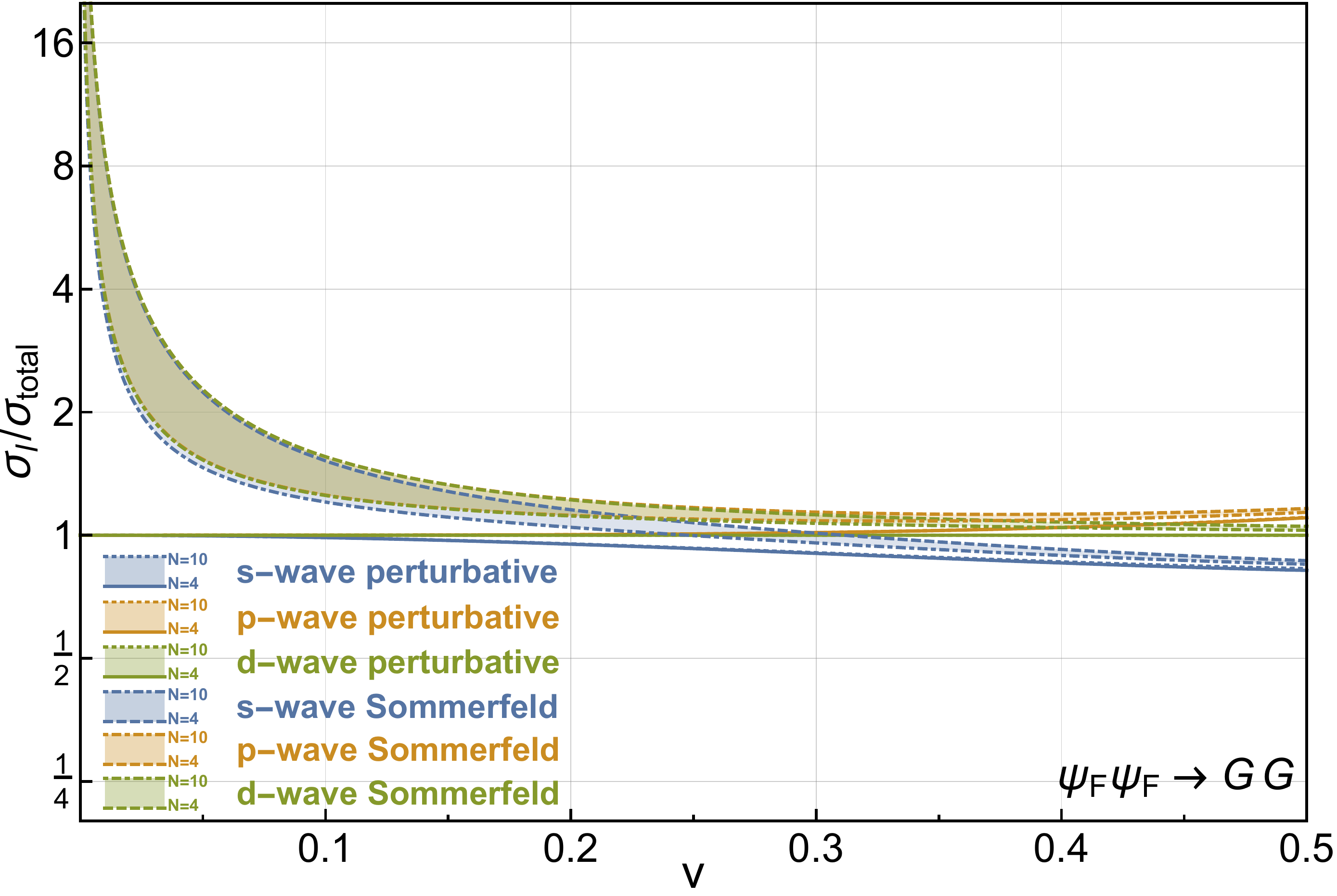}
	\includegraphics[width=0.495\textwidth]{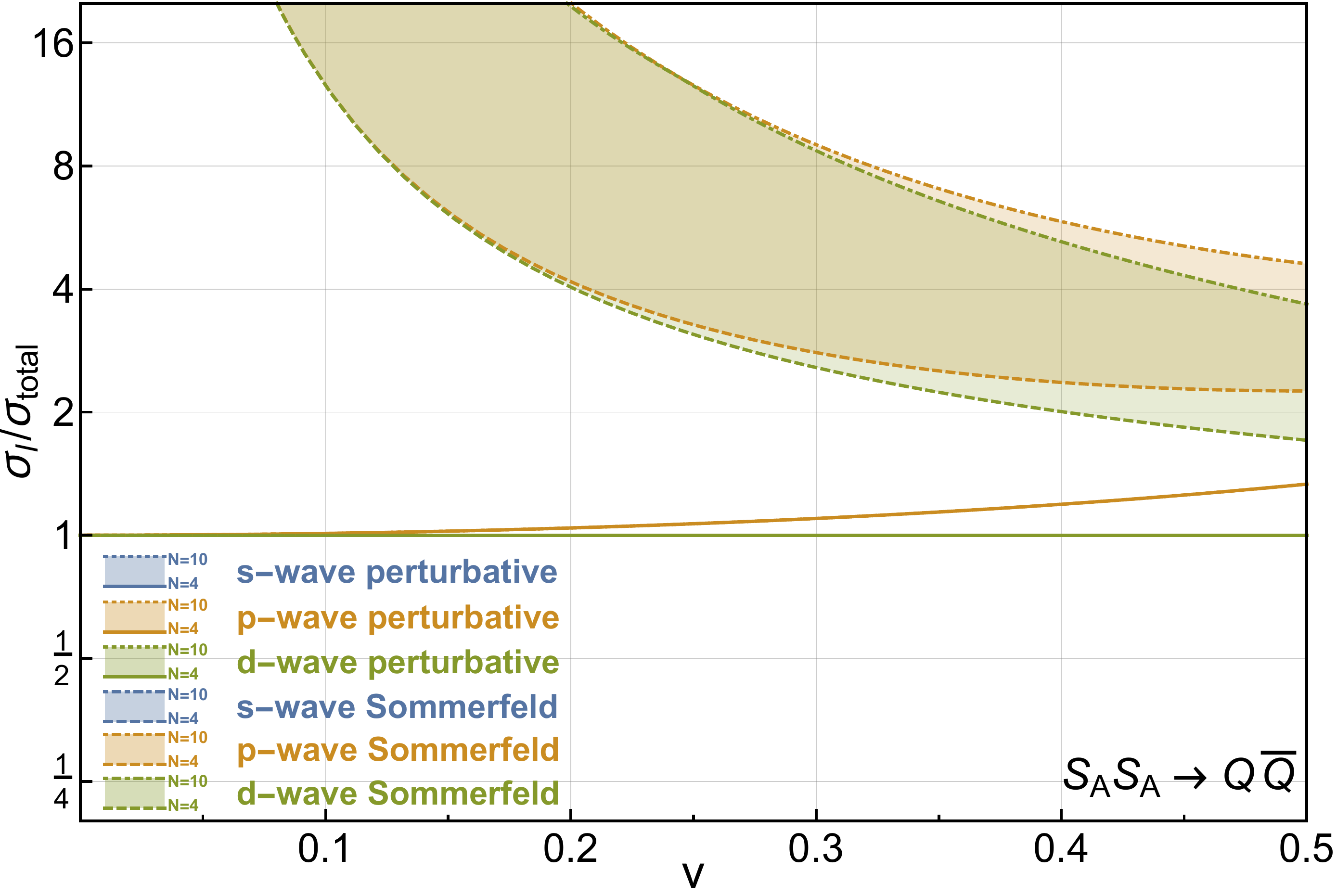}
	\includegraphics[width=0.495\textwidth]{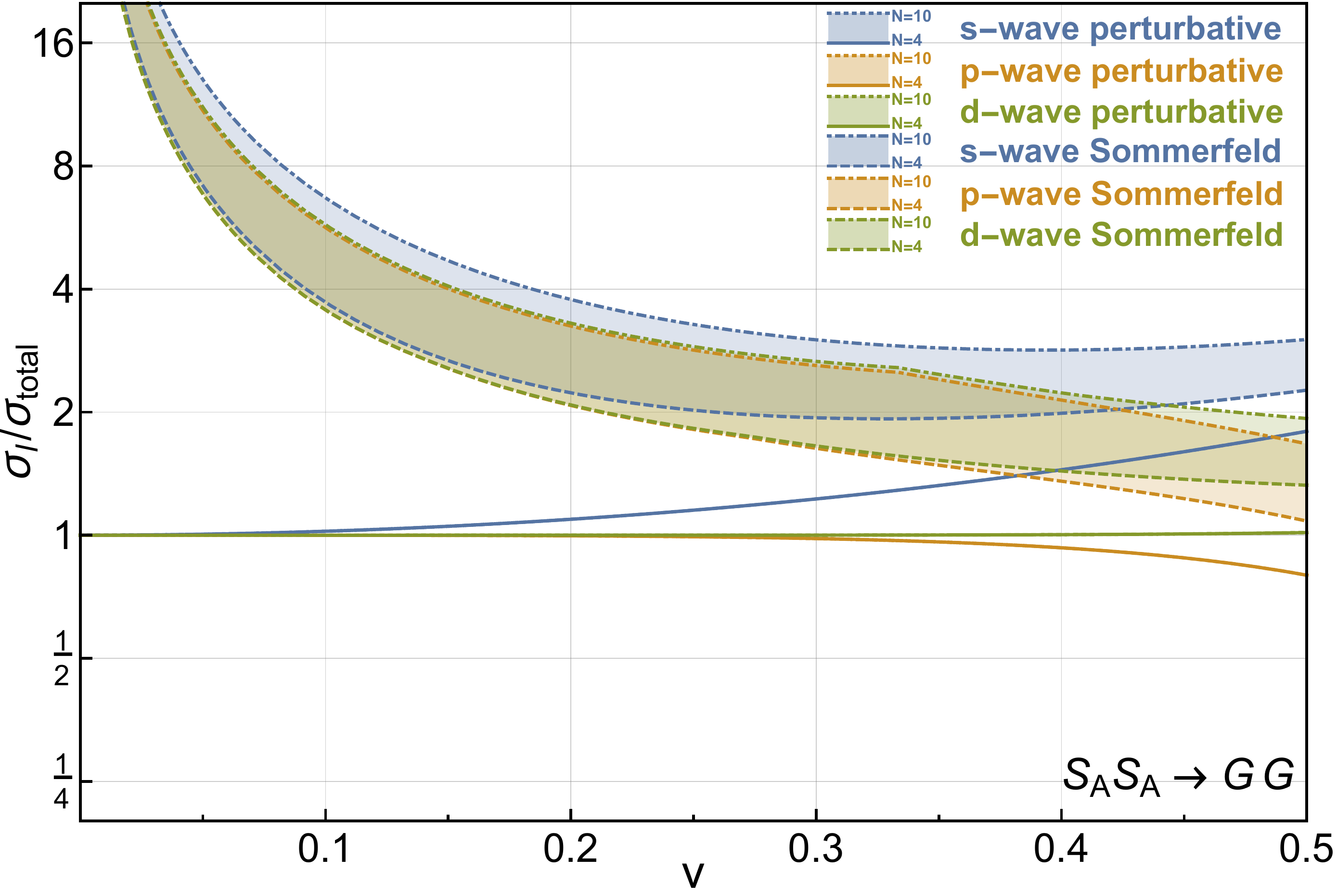}
	\includegraphics[width=0.495\textwidth]{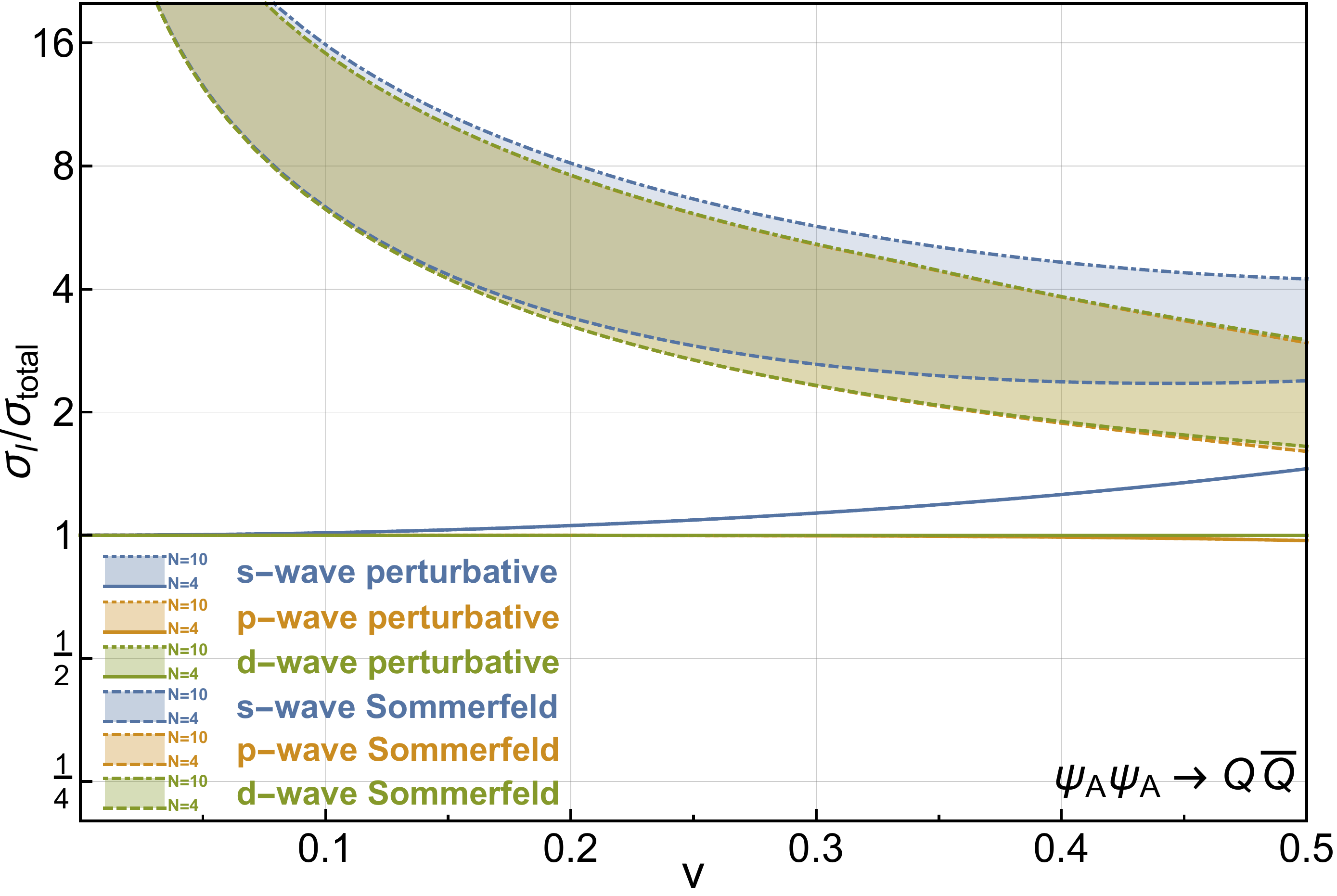}
	\includegraphics[width=0.495\textwidth]{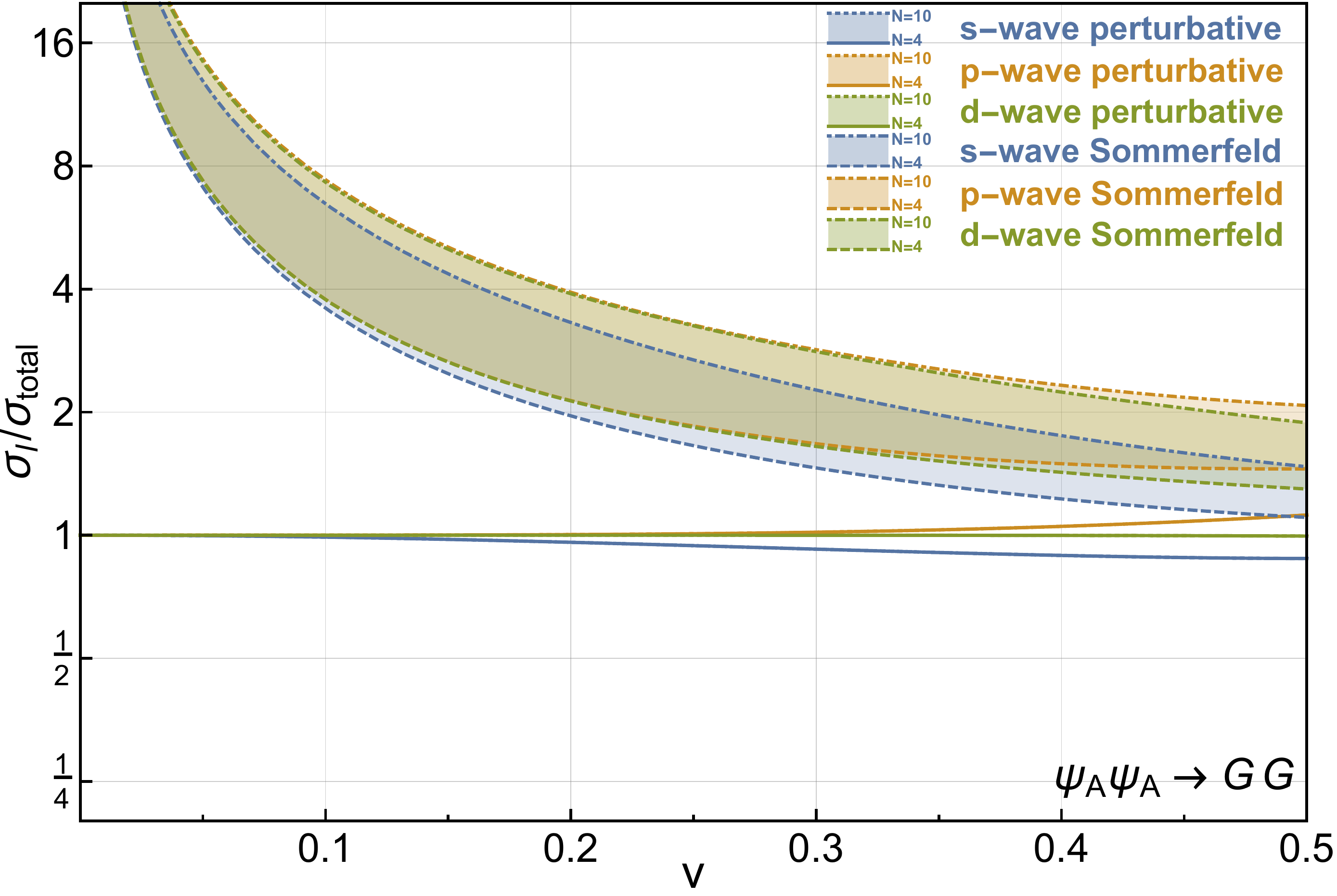}
	\caption{Ratios of the perturbative (solid lines) and Sommerfeld-corrected cross sections (dashed lines) expanded up to the $s$-wave (blue), $p$-wave (orange) and $d$-wave (green) over the total perturbative cross section. We show the ratios as a band corresponding to $4 \leq N \leq 10$ for a specific process and representation (either $\irrep{F}$ or $\irrep{A}$), which is denoted by the subscript on the $\Phi$ fields.}
	\label{fig:annihilation:cross:section:sun}
\end{figure}

We now use the results from section~\ref{sec:sun:color:decomposition} as well as the methodology described in section~\ref{sec:decomposing:qcd:sommerfeld} to derive the Sommerfeld correction factors for the $\Phi \, \overline{\Phi} \to Q_i \, \bar{Q}_j$ and the  $\Phi \, \overline{\Phi} \to G^a \, G^b$ annihilation processes. For general $N$, these factors now read
\begin{equation} \label{eq:sun:sommerfeld:corrections}
	\begin{aligned}
		\sigma^{(S)}_{\irrep{F} \otimes \overline{\irrep{F}} \to Q\, \overline{Q}} & = \sigma^{(S)}_C \left[ - \frac{\alpha_N}{2N} \right] \\
		\sigma^{(S)}_{\irrep{F} \otimes \overline{\irrep{F}} \to G \, G} & = \begin{cases} \frac{2}{N^2 - 2} \sigma^{(S)}_C \left[ \frac{(N^2 - 1) \alpha_N}{2N} \right] + \frac{N^2 - 4}{N^2 - 2} \sigma^{(S)}_C \left[ - \frac{\alpha_N}{2N} \right] & \mathrm{even} \,\, l + s \\ \sigma^{(S)}_C \left[ - \frac{\alpha_N}{2N} \right] & \phantom{!} \mathrm{odd} \,\, l + s \end{cases} \\
		\sigma^{(S)}_{\irrep{A} \otimes \irrep{A} \to Q \, \overline{Q}} & = \sigma^{(S)}_C \left[ \frac{N \alpha_N}{2} \right] \\
		\sigma^{(S)}_{\irrep{A} \otimes \irrep{A} \to G \, G} & = \begin{cases} \frac{4}{3 (N^2 - 1)} \sigma^{(S)}_C \left[ N \alpha_N \right] + \frac{1}{3} \sigma^{(S)}_C \left[ \frac{N \alpha_N}{2} \right] & \\ + \frac{N-3}{3(N - 1)} \sigma^{(S)}_C\left[ \alpha_N \right] + \frac{N+3}{3(N + 1)} \sigma^{(S)}_C\left[ - \alpha_N \right] & \mathrm{even} \,\, l + s \\ \sigma^{(S)}_C \left[ \frac{N \alpha_N}{2} \right] & \phantom{!} \mathrm{odd} \,\, l + s \end{cases} .
	\end{aligned}
\end{equation}
The ratios of the $s$, $p$ and $d$-wave annihilation cross sections are shown in figure~\ref{fig:annihilation:cross:section:sun} for the $\Phi \, \overline{\Phi} \to Q_i \, \bar{Q}_j$ and $\Phi \, \overline{\Phi} \to G^a \, G^b$ processes with $\Phi$ being either a scalar or a fermion, in either the fundamental or the adjoint representation of $SU(N)$. As in figure~\ref{fig:annihilation:cross:section}, we consider velocity expansions of the cross section up to the $s$-wave, the $p$-wave and the $d$-wave but this time, we show the values of these different cross sections for $4 \leq N \leq 10$. Contrary to the $SU(3)$ case, we do not evaluate $\alpha_N$ at the scale of the momenta of the incoming particles and instead set the coupling entering into the Sommerfeld corrections to be $\alpha_N^\mathrm{Sommerfeld} (\hat{\mu}) = 0.1$. For the typical momenta considered here, this value is lower than the ones encountered in the QCD case, thereby leading to conservative estimates of the Sommerfeld effect in strongly coupled theories.

\afterpage{\clearpage}

For fundamental particles in the initial state, the Sommerfeld corrections become negligible in the large $N$ limit. This result can be understood by noting that, in equation~\eqref{eq:sun:sommerfeld:corrections}, either the effective couplings for the Coulomb potentials or the coefficients of the $\sigma_C$ cross sections are inversely proportional to powers of $N$. For initial state particles in the adjoint representation, however, the dominant contributions in the large $N$ limit arise from terms of the form $\sigma_C\left[\frac{N\alpha_N}{2}\right]$. In this case, the Sommerfeld enhancement will therefore grow with $N$ for each partial wave contribution, as can be observed in figure~\ref{fig:annihilation:cross:section:sun}. Note that in this scenario, the Sommerfeld enhancement is extremely relevant at typical freeze-out velocities and taking it into account is essential for relic abundance computations.

\subsection{Messenger particles}
\label{sec:sun:messenger:particles}
One particular scenario often encountered in the literature is the existence of new particles that are charged both under QCD and under a new $SU(N)$ gauge group. These particles notably play the roles of messengers between the Standard Model and the dark sector in hidden valley models~\cite{Juknevich:2009ji,Juknevich:2009gg,Soni:2017nlm}. In this case the non-relativistic potential for the Sommerfeld effect is the sum of the $SU(3)$ and the $SU(N)$ potential and Sommerfeld correction factors are modified accordingly~\cite{Zhang:2013qza}. The total potential is given as
\begin{equation} \label{eq:potential:double:charge}
	V = \frac{\alpha_s (\hat{\mu})}{r} \sum_a T^a_{\irrep{R}} \otimes T^a_{\irrep{R}'} + \frac{\alpha_N (\hat{\mu})}{r} \sum_a T^a_{\irrep{P}} \otimes T^a_{\irrep{P}'} \, ,
\end{equation}
where $\irrep{R}$ is a representation of $SU(3)$ and $\irrep{P}$ is a representation of $SU(N)$. Computing this potential for initial states in different $SU(3)\times SU(N)$ representations can be done by applying equation~\eqref{eq:potential:color:decomposition} to each of the terms on the right hand side of equation~\eqref{eq:potential:double:charge} separately.  For a given annihilation process, the Clebsch-Gordan coefficient for an initial state with given $SU(3)\times SU(N)$ quantum numbers is the product of the coefficients corresponding to the $SU(3)$ and the $SU(N)$ representations. These coefficients can be readily computed using equations~\eqref{eq:decomposition:sun:quarks:odd}, \eqref{eq:decomposition:sun:fundamentals:even} and \eqref{eq:decomposition:sun:adjoints:even}. In what follows, we will apply this procedure to the particular case of particles charged under the fundamental representations of both $SU(3)$ and $SU(N)$.

\subsection{Application: bifundamental messengers}
\label{sec:sun:applications}
In models where the gauge bosons of the dark $SU(N)$ either are dark radiation or form dark glueballs a connection between the dark sector and the Standard Model needs to exist to ensure thermal equilibrium. This connection can be established by introducing messenger particles charged both under QCD and under the dark $SU(N)$ gauge group. These particles are initially in thermal equilibrium with the SM and therefore annihilate to SM particles until they freeze out. When the temperature of the universe drops below the confining scale of the theory at later times, these messengers form bound states that decay to dark gauge bosons that ultimately form stable glueball dark matter candidates. The strength of the messenger annihilation channels to the visible and dark sectors will therefore set the dark matter relic abundance. 

\begin{figure}[!ht]
	\centering
	\includegraphics[width=0.8\textwidth]{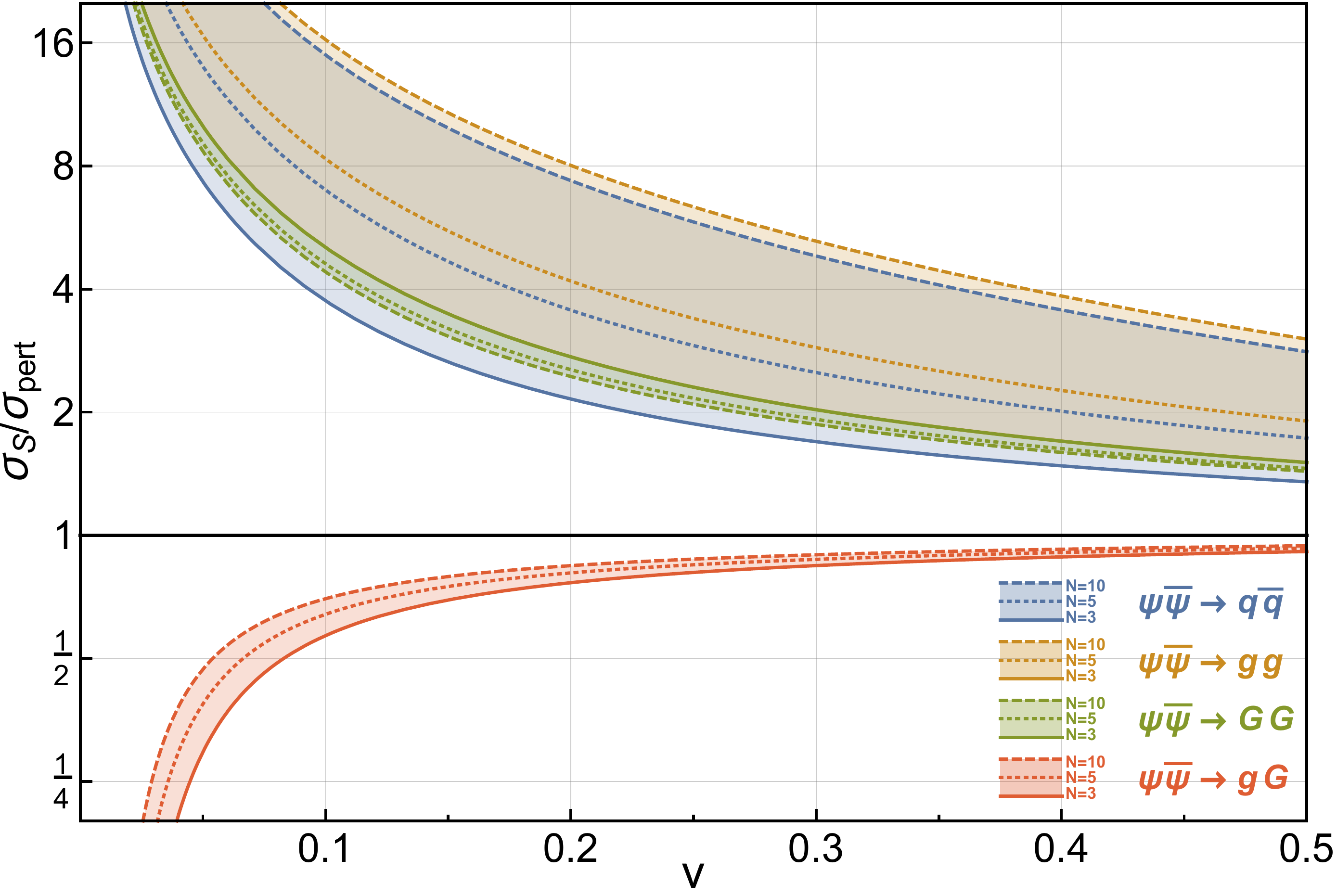}
	\caption{Sommerfeld correction factors for the $s$-wave annihilation cross sections of fermionic messengers in the fundamental representation of both QCD and $SU(N)$. The different colors show the relevant annihilation processes and the different lines represent $N=3$ (solid), $N=5$ (dotted) and $N=10$ (dashed).}
	\label{fig:annihilation:cross:section:model}
\end{figure}

In what follows, we consider a fermionic messenger particle $\psi$ charged as a triplet under QCD and as a fundamental under $SU(N)$. In this scenario, $\psi$ can annihilate either to $g \, g$, $q \, \bar{q}$, $G \, G$ or $g \, G$, where $G$ is the massless dark gauge boson for the $SU(N)$ gauge group. The first two processes occur through the QCD interaction and, since the final states are $SU(N)$ singlets, the initial $\psi \, \bar{\psi}$ state must also be an $SU(N)$ singlet. The different QCD representations for $\psi \, \bar{\psi}$ as well as their corresponding Clebsch-Gordan coefficients are therefore the ones derived in section~\ref{sec:decomposing:qcd:sommerfeld}.  As outlined in section~\ref{sec:sun:messenger:particles}, however, the non-relativistic potential between the two initial state particles will now have an additional term corresponding to the exchange of dark gluons. Since $\psi \, \bar{\psi}$ has to be an $SU(N)$ singlet, the new potential will be of the form
\begin{equation}
	V = V_{SU(3)} - \frac{N^2 - 1}{2 N} \frac{\alpha_N}{r} ,
\end{equation}
where $V_{SU(3)}$ is given in equation~\eqref{eq:decomposed:potentials}. The $\psi \, \bar{\psi} \to G \, G$ process occurs through $SU(N)$ interactions and has been studied in section~\ref{sec:sun:sommerfeld:corrections}. The results from this section can be directly applied to this scenario with the potential being modified as
\begin{equation}
	V = - \frac{4}{3} \frac{\alpha_s}{r} + V_{SU(N)} ,
\end{equation}
where $V_{SU(N)}$ is given in equation~\eqref{eq:sun:product:decomposition}. Note here that $\psi \, \bar{\psi}$ now is an $SU(3)$ singlet.

Finally, the $\psi \, \bar{\psi} \to g \, G$ process has not been studied before and has been not been taken into account in~\cite{Soni:2017nlm}. For this annihilation channel, gauge conservation constrains the $\psi \, \bar{\psi}$ initial state to be in the adjoint representation of both $SU(3)$ and $SU(N)$. Hence, there is no need to compute any Clebsch-Gordan coefficient and the potential will now read
\begin{equation}
	V = \frac{1}{6} \frac{\alpha_s}{r} + \frac{1}{2 N} \frac{\alpha_N}{r} .
\end{equation}
The Sommerfeld-corrected annihilation cross sections for all these processes in the $s$-wave can then be expressed as
\begin{equation}
	\begin{aligned}
		\left( \sigma v \right)_{\psi \, \bar{\psi} \to q \, \bar{q}} = & \, 6 \times \frac{\pi \alpha_s^2}{9 N m_Q^2} \times S \left( - \frac{\alpha_s}{6 \beta} + \frac{N^2 \! - \! 1}{2 N} \frac{\alpha_N}{\beta} \right) \\
		\left( \sigma v \right)_{\psi \, \bar{\psi} \to g \, g} = & \, \frac{7 \pi \alpha_s^2}{54 N m_Q^2} \times \left[ \frac{2}{7} S \left( \frac{4 \alpha_s}{3 \beta} + \frac{N^2 \! - \! 1}{2 N} \frac{\alpha_N}{\beta} \right) + \frac{5}{7} S \left( - \frac{\alpha_s}{6 \beta} + \frac{N^2 \! - \! 1}{2 N} \frac{\alpha_N}{\beta} \right) \right] \\
		\left( \sigma v \right)_{\psi \, \bar{\psi} \to G \, G} = & \, \frac{(N^2 \! - \! 1)(N^2 \! - \! 2) \pi \alpha_N^2}{48 N^3 m_Q^2} \\
		& \, \times \left[ \frac{2}{N^2 \! - \! 2} S \left( \frac{4 \alpha_s}{3 \beta} + \frac{N^2 \! - \! 1}{2 N} \frac{\alpha_N}{\beta} \right) + \frac{N^2 \! - \! 4}{N^2 \! - \! 2} S \left( \frac{4 \alpha_s}{3 \beta} - \frac{1}{2N} \frac{\alpha_N}{\beta} \right) \right] \\
		\left( \sigma v \right)_{\psi \, \bar{\psi} \to g \, G} = & \, \frac{2 (N^2 \! - \! 1) \pi \alpha_s \alpha_N}{9 N^2 m_Q^2} \times S \left( - \frac{\alpha_s}{6 \beta} - \frac{1}{2N} \frac{\alpha_N}{\beta} \right) ,
	\end{aligned}
\end{equation}
where the Sommerfeld factor $S(x)$ is given in equation~\eqref{eq:sommerfeld:corrections:lowestorder}. The ratios of these cross sections over the $s$-wave perturbative cross sections for each process are shown in figure~\ref{fig:annihilation:cross:section:model} for $N = 3, 5, 10$ and with $\alpha_s(\hat{\mu}) = \alpha_N(\hat{\mu}) = 0.1$. For typical freeze-out velocities $v \sim 0.2$, the Sommerfeld effect can lead to a factor of two to eight enhancement of the annihilation cross section for most processes. Although this enhancement could be slightly mitigated by the reduction of the cross section for $\psi \, \bar{\psi} \to g \, G$, this reduction is in general much less pronounced than the enhancement observed for the other processes, especially as $N$ increases. Taking the Sommerfeld corrections into account for the annihilation of messenger particles is therefore essential to derive robust cosmological bounds for the hidden sector models of dark matter discussed in~\cite{Soni:2017nlm,Feng:2011ik,Boddy:2014qxa,Boddy:2014yra,Harigaya:2016nlg}.

\section{Conclusions}
\label{sec:conclusions}
In this work we have derived analytical expressions for the Sommerfeld corrections of the annihilation of colored particles. These expressions result from combining two orthogonal procedures: deriving Sommerfeld corrections for partial waves beyond the leading order and decomposing the QCD potential into Coulomb potentials. Our results significantly improve on existing literature and allow to combine higher-order velocity corrections with the QCD nature of these annihilation processes. These analytical expressions can readily be applied to any type of annihilation of colored particles in dark sector. The only necessary step is to expand the annihilation cross sections into states of definite orbital angular momentum and spin $(l, s)$ and then apply the correction factors as presented in our work.

For consistently applying Sommerfeld correction factors for higher partial waves we showed it is necessary to expand the annihilation amplitudes in $(l, s)$ states. Then one can further expand these states in powers of the momentum and solve the non-relativistic Schrödinger equation for each of the states separately. From these solutions one obtains the analytic Sommerfeld-correction factors for all orders in the partial wave expansion and all  powers of the momentum. We express these results conveniently as the Sommerferld-correction factor for the $s$-wave times an analytic distortion factor specific to each term in the partial wave expansion.

The QCD nature of the process poses a challenge for the analytic calculation of the Sommerfeld corrections which can be overcome by decomposing the potential into a linear combination of Coulomb potentials. This procedure, however, depends crucially on the symmetry properties of the amplitude. With an expansion of the amplitude in $(l, s)$ states these properties become apparent. The color structure then simplifies and becomes independent of the kinematics of the process. Then the color-dependent part of the annihilation amplitude can be treated separately and later combined with the Sommerfeld corrections for the partial wave components. 

Finally, we apply these results to several colored dark sectors with a singlet dark matter candidate, where the annihilation of the colored states is solely responsible for setting the relic abundance. We show that for particles of any spin --- scalar, fermion, vector ---  and in the triplet, sextet or octet representation of QCD Sommerfeld corrections are sizable. A consistent and precise inclusion of these effects is therefore essential in understanding the specific details of a possible colored dark sector. In an accompanying paper we present the full study of several types of colored dark sectors where we include precise determination of the relic density and discuss the full phenomenology of these models.

We also present the first calculation of the Sommerfeld corrections for dark sectors charged under general $SU(N)$ gauge groups. These corrections are especially relevant in scenarios where confinement occurs after freeze-out, since the new gauge group remains unbroken and the gauge coupling is sizable. Although non-perturbative effects for these models have been previously overlooked in the literature, we showed that the Sommerfeld corrections can drastically modify the annihilation cross section of dark sector particles, and therefore the dark matter relic density. We advocate for taking these corrections into account in future in-depth studies of these models.

We conclude by emphasizing that the procedure described in this work is not restricted to the annihilation of identical particles. Notably, our method also applies to processes like the annihilation of a triplet and an octet of QCD --- for example squark-gluino annihilation in supersymmetry. Henceforth, Sommerfeld corrections for models with extended dark sectors and multiple gauge groups can be easily included using the presented formalism.

\acknowledgments
We would like to thank Andrzej Hryczuk, Felix Yu and Jos\'{e} Zurita for valuable discussions. This research is supported by the Cluster of Excellence Precision Physics, Fundamental Interactions and Structure of Matter (PRISMA-EXC 1098), by the ERC Advanced Grant EFT4LHC of the European Research Council, and by the Mainz Institute for Theoretical Physics.

\appendix
\section{Partial wave cross sections}
\label{sec:xsec:partial:wave}
This appendix details the conventions used to compute the annihilation cross sections in the \texttt{Mathematica} notebook attached to this paper~\cite{ElHedri:2016pac}. In order for the amplitudes to be of the form of equation~\eqref{eq:amplitude}, we work in the so-called \emph{final frame} where the momenta of the final state are along the $z$-axis and the momenta of the initial states are characterized by the angles $\theta$ and $\phi$. In order for the polarization vectors of the gluons and of the vector $\Phi$ to be well-defined, we compute the amplitudes in the helicity basis based on~\cite{Just:2008xy}. In this basis, for the annihilation processes we consider in the main body of the paper the $\phi$ dependence of the amplitude is well-known 
\begin{equation}
	\mathcal{A}_{\lambda_1, \lambda_2, \lambda_3, \lambda_4}(p, \theta, \phi) \equiv \mathcal{A}_{\lambda_1, \lambda_2, \lambda_3, \lambda_4}(p, \theta, \phi = 0) e^{i(\lambda_i - \lambda_f)\phi} ,
\end{equation}
with $\lambda_i = \lambda_1 - \lambda_2$ and $\lambda_f = \lambda_3 - \lambda_4$.\footnote{For other processes like the ones discussed in appendices~\ref{sec:exotic:qcd:triplet:octet} and~\ref{sec:exotic:qcd:triplet:triplet} one needs to take into account the full $\phi$-dependence of the amplitude.} For these processes, we therefore compute the amplitudes for $\phi = 0$ and inject the azimuthal phase factor into the final expression. 

Assuming the quarks to be massless, the momenta for the annihilation of a pair of $\Phi$'s into quark and gluon pairs are
\begin{equation} \label{eq:momenta}
	\begin{aligned}
		p_1 & = (E, p \sin \theta \cos \phi, p \sin \theta \sin \phi, p \cos \theta) \\
		p_2 & = (E, -p \sin \theta \cos \phi, -p \sin \theta \sin \phi, -p \cos \theta) \\
		p_3 & = (E, 0, 0, E) \\
		p_4 & = (E, 0, 0, -E) ,
	\end{aligned}
\end{equation}
with $E = \sqrt{p^2 + m^2}$. The spinors for a particle of helicity $\pm \frac{1}{2}$ and mass $m$ moving in the direction $(\theta, \phi)$ are
\begin{equation} \label{eq:spinors:rotated}
	\renewcommand\arraystretch{1}
	\begin{aligned}
		u_+ (p, m, \theta, \phi) & = \mathcal{R} (\theta, \phi) \cdot
		\begin{pmatrix}
			\sqrt{E - p} \\
			0 \\
			\sqrt{E + p} \\
			0
		\end{pmatrix}
		\quad\quad
		u_- (p, m, \theta, \phi) = \mathcal{R} (\theta, \phi) \cdot
		\begin{pmatrix}
			0 \\
			\sqrt{E + p} \\
			0 \\
			\sqrt{E - p}
		\end{pmatrix} \\
		v_+ (p, m, \theta, \phi) &= \mathcal{R} (\theta, \phi) \cdot
		\begin{pmatrix}
			0 \\
			-\sqrt{E + p} \\
			0 \\
			\sqrt{E - p}
		\end{pmatrix}
		\quad
		v_- (p, m, \theta, \phi) = \mathcal{R} (\theta, \phi) \cdot
		\begin{pmatrix}
			\sqrt{E - p} \\
			0 \\
			-\sqrt{E + p} \\
			0
		\end{pmatrix} ,
	\end{aligned}
\end{equation}
with  
\begin{equation}
	\mathcal{R} (\theta, \phi) =
	\begin{pmatrix}
		\cos \frac{\theta}{2} & -\sin \frac{\theta}{2} e^{-i \phi} & 0 & 0 \\ 
		\sin \frac{\theta}{2} e^{i \phi} & \cos \frac{\theta}{2} & 0 & 0 \\ 
		0 & 0 & \cos \frac{\theta}{2} & -\sin \frac{\theta}{2} e^{-i \phi} \\ 
		0 & 0 & \sin \frac{\theta}{2} e^{i \phi} & \cos \frac{\theta}{2}
	\end{pmatrix} .
\end{equation}
The spinors for a particle moving in the opposite direction are obtained in~\cite{Just:2008xy} as well. They are very similar to the expressions in equation~\eqref{eq:spinors:rotated} and are given in the attached \texttt{Mathematica} notebook~\cite{ElHedri:2016pac}. When computing amplitudes involving fermion currents, we define the gamma matrices in the Weyl basis.

The transverse polarization vectors corresponding to a final state gluon of momentum $p_3$ or to a vector $\Phi$ of momentum $p_1$ are
\begin{equation}
	\epsilon_\pm^{(1)}(\theta) = \frac{1}{\sqrt{2}} \left( 0, \mp \cos \theta, - i, \pm \sin \theta \right) ,
\end{equation}
while the longitudinal polarization vector corresponding to a vector $\Phi$ of momentum $p_1$ is
\begin{equation}
	\epsilon_0^{(1)}(p, m, \theta) = \left( \frac{p}{m}, \frac{E}{m} \sin \theta, 0, \frac{E}{m} \cos \theta \right) .
\end{equation}
Similarly, the transverse polarization vectors corresponding to a final state gluon of momentum $p_4$ or to a vector $\Phi$ of momentum $p_2$ are
\begin{equation}
	\epsilon_\pm^{(2)}(\theta) = \frac{1}{\sqrt{2}} \left(0, \pm \cos \theta, - i, \mp \sin \theta \right) ,
\end{equation}
while the longitudinal polarization vector corresponding to a vector $\Phi$ of momentum $p_2$ is
\begin{equation}
	\epsilon_0^{(2)}(p, m, \theta) = \left(-\frac{p}{m}, \frac{E}{m} \sin \theta, 0, \frac{E}{m} \cos \theta \right) .
\end{equation}
The full $\phi$-dependence for polarization vectors can be found in the attached \texttt{Mathematica} notebook~\cite{ElHedri:2016pac}. We use~\cite{Chung:1971ri} to convert the helicity amplitudes into amplitudes for definite spin states $m_1, m_2, m_3, m_4$ (with $m_i$ being the $z$-component of the spin of particle $i$) using
\begin{equation}
	\mathcal{A}_{m_1, m_2, m_3, m_4} (p, \theta, \phi) \! = \! \! \sum_{\lambda_1, \lambda_2} \! D^{s_1\,*}_{m_1, \lambda_1} \! (\phi, \theta, -\phi) D^{s_2\,*}_{m_2, -\lambda_2} \! (\phi, \theta, -\phi) \mathcal{A}_{\lambda_1, \lambda_2, m_3, -m_4}(p, \theta, \phi)
\end{equation}
where the $D^s_{m,\lambda}$ are the Wigner D-functions. In this formula, we used $\lambda_3 = m_3$ and $\lambda_4 = -m_4$ for final states, whose momenta are along the $z$-axis.

\section{Color decomposition}
\label{sec:color:decomposition}
In this appendix we describe the decomposition of the group structure of the amplitudes discussed in section~\ref{sec:decomposing:qcd:cross:section} for QCD and in section~\ref{sec:sun:color:decomposition} for $SU(N)$. Moreover, later in this appendix we discuss the decomposition and Sommerfeld corrections for more exotic particles present in colored dark sectors.

\subsection{Amplitude tensor decomposition}
\label{sec:color:decomposition:amplitude}
In this section we describe how to decompose a colored amplitude into several channels of definite color. Then using these expressions and using a specific form for the color part of the amplitude as obtained in section~\ref{sec:decomposing:qcd:cross:section} we square the amplitude and find the decomposition of the total cross section. In principle one can decompose amplitudes which may be any product of representations of $SU(N)$, however, here we restrict ourselves to $\irrep{R} \otimes \overline{\irrep{R}}$ with $\irrep{R} = \irrep{F}$ (fundamental), $\irrep{S}$ ($\frac{N(N+1)}{2}$-dimensional symmetric) and $\irrep{A}$ (adjoint). The decomposition for these channels is given in equation~\eqref{eq:representation:product} for QCD and in equation~\eqref{eq:sun:product:decomposition} for $SU(N)$. More exotic combinations are discussed for QCD in the next two sections of this appendix. To decompose the amplitudes we base ourself on the method of tensor decomposition as described in~\cite{Georgi:1999wka} and use fundamental indices for all representations. To switch between $(T_{\irrep{R}}^a)^i_j$ where $i,j$ run from $1$ to $d_{\irrep{R}}$ and the fundamental indices one can use the Clebsch-Gordan coefficients of the representation with respect to fundamentals of $SU(N)$. We focus purely on the color of $\Phi$  in $A$ and drop the color dependence of the remaining part in the amplitudes. This method has been put forward already in~\cite{deSimone:2014pda} for the triplet and the octet in QCD and we extend these results to arbitrary $N$. Parts of these calculation have been done using \texttt{LieArt}~\cite{Feger:2012bs} and \texttt{ColorMath}~\cite{Sjodahl:2012nk}. For the product of two fundamentals we can write the tensor product as $A^k_i = v^i w_j$ which contains the full color dependence of the total amplitude. We split up this part of the amplitude as
\begin{equation} \label{eq:decomposition:color:fundamental}
	\begin{aligned}
		A^i_j & = [\irrep{1}]^i_j + [\irrep{A}]^i_j \\
		[\irrep{1}]^i_j & = \frac{1}{N} \delta^i_j A^m_m \\
		[\irrep{A}]^i_j & = A^i_j - \frac{1}{N} \delta^i_j A^m_m .
	\end{aligned}
\end{equation}
In here the indices $i, j = 1, \cdots, N$ represent the color of the $\Phi_{i,j}$. For the product of two symmetric representations $\irrep{S}$ the situation is slightly more complicated as one needs to represent each $\Phi_u$ where $u = 1, \cdots, \frac{1}{2}N(N+1)$ with two fundamental indices $i, j = 1, \cdots, N$. We can now write $A^{ij}_{kl} = v^{ij} w_{kl}$, which now has to be symmetric under the transformations $i \leftrightarrow j$, $k \leftrightarrow l$. Transforming between both representations can be done using the respective Clebsch-Gordan coefficients~\cite{Han:2009ya}. The symmetricity representation decomposes as
\begin{equation} \label{eq:decomposition:color:symmetric}
	\begin{aligned}
		A^{ij}_{kl} & = [\irrep{1}]^{ij}_{kl} + [\irrep{A}]^{ij}_{kl} + [\irrep{D}]^{ij}_{kl} \\
		[\irrep{1}]^{ij}_{kl} & = \frac{1}{N (N \! + \! 1)} A^{mn}_{mn} \left( \delta^i_k \delta^j_l \! + \! \delta^i_l \delta^j_k \right) \\
		[\irrep{A}]^{ij}_{kl} & = \frac{1}{N \! + \! 2} \left[ \delta^i_k A^{mj}_{ml} \! + \! \delta^j_k A^{mi}_{ml} \! + \! \delta^j_l A^{mi}_{mk} \! + \! \delta^i_l A^{mj}_{mk}\right] \! - \! \frac{2}{N (N \! + \! 2)} A^{mn}_{mn} \left( \delta^i_k \delta^j_l \! + \! \delta^i_l \delta^j_k \right) \\
		[\irrep{D}]^{ij}_{kl} & = A^{ij}_{kl} \! - \! \frac{1}{N \! + \! 2} \left[ \delta^i_k A^{mj}_{ml} \! + \! \delta^j_k A^{mi}_{ml} \! + \! \delta^j_l A^{mi}_{mk} \! + \! \delta^i_l A^{mj}_{mk}\right] \! + \! \frac{1}{(N \! + \! 1)(N \! + \! 2)} A^{mn}_{mn} \left( \delta^i_k \delta^j_l \! + \! \delta^i_l \delta^j_k \right) .
	\end{aligned}
\end{equation}
For the adjoint we can write $A^{ij}_{kl} = v^i_k w^j_l$ with $A^{mj}_{ml} = A^{im}_{km} = 0$. Again we can transform to adjoint indices $a = 1, \cdots, N^2 - 1$ by using the respective Clebsch-Gordan coefficients which are obtained directly from the generators of $SU(N)$. The adjoint decomposes as
\begin{equation} \label{eq:decomposition:color:adjoint}
	\begin{aligned}
		A^{ij}_{kl} = & [\irrep{1}_\textbf{S}]^{ij}_{kl} + [\irrep{A}_\textbf{A}]^{ij}_{kl} + [\irrep{A}_\textbf{S}]^{ij}_{kl} + [\irrep{B}_\textbf{S}]^{ij}_{kl} + [\irrep{C}_\textbf{A}]^{ij}_{kl} + [\overline{\irrep{C}}_\textbf{A}]^{ij}_{kl} + [\irrep{D}_\textbf{S}]^{ij}_{kl} \\
		[\irrep{1}_\textbf{S}]^{ij}_{kl} = & \frac{1}{N \! - \! N^3} A^{mn}_{nm} \left(\delta^i_k \delta^j_l - N \delta^i_l \delta^j_k\right) \\
		[\irrep{A}_\textbf{A}]^{ij}_{kl} = & \frac{1}{2 N} \left[ - \delta^i_l (A^{jm}_{mk} - A^{mj}_{km}) + \delta^j_k (A^{im}_{ml} - A^{mi}_{lm}) \right] \\
		[\irrep{A}_\textbf{S}]^{ij}_{kl} = & \frac{4}{N(N^2 \! - \! 4)} A^{mn}_{nm} \left(\delta^i_k\delta^j_l - \frac{N}{2} \delta^i_l\delta^j_k\right) \\
		& + \! \frac{1}{4 \! - \! N^2} \! \left[ \delta^i_k (A^{jm}_{ml} \! + \! A^{mj}_{lm}) \! - \! \frac{N}{2} \delta^i_l (A^{jm}_{mk} \! + \! A^{mj}_{km}) \! - \! \frac{N}{2} \delta^j_k (A^{im}_{ml} \! + \! A^{mi}_{lm}) \! + \! \delta^j_l (A^{im}_{mk} \! + \! A^{mi}_{km}) \right] \\
		[\irrep{B}_\textbf{S}]^{ij}_{kl} = & \frac{1}{4} (A^{ij}_{kl} - A^{ji}_{kl} - A^{ij}_{lk} + A^{ji}_{lk}) - \frac{1}{2 (N^2 \! - \! 3N \! + \! 2)} A^{mn}_{nm} \left(\delta^i_k\delta^j_l - \delta^i_l\delta^j_k\right) \\
		& + \! \frac{1}{4(N \! - \! 2)} \! \left[ \delta^i_k (A^{jm}_{ml} \! + \! A^{mj}_{lm}) \! - \! \delta^i_l (A^{jm}_{mk} \! + \! A^{mj}_{km}) \! - \! \delta^j_k (A^{im}_{ml} \! + \! A^{mi}_{lm}) \! + \! \delta^j_l (A^{im}_{mk} \! + \! A^{mi}_{km}) \right] \\
		[\irrep{C}_\textbf{A}]^{ij}_{kl} = & \frac{1}{4} (A^{ij}_{kl} + A^{ji}_{kl} - A^{ij}_{lk} - A^{ji}_{lk}) \\
		& - \! \frac{1}{4N} \! \left[ \delta^i_k (A^{jm}_{ml} \! - \! A^{mj}_{lm}) \! - \! \delta^i_l (A^{jm}_{mk} \! - \! A^{mj}_{km}) \! + \! \delta^j_k (A^{im}_{ml} \! - \! A^{mi}_{lm}) \! - \! \delta^j_l (A^{im}_{mk} \! - \! A^{mi}_{km}) \right] \\
		[\overline{\irrep{C}}_\textbf{A}]^{ij}_{kl} = & \frac{1}{4} (A^{ij}_{kl} - A^{ji}_{kl} + A^{ij}_{lk} - A^{ji}_{lk}) \\
		& + \! \frac{1}{4N} \! \left[ \delta^i_k (A^{jm}_{ml} \! - \! A^{mj}_{lm}) \! + \! \delta^i_l (A^{jm}_{mk} \! - \! A^{mj}_{km}) \! - \! \delta^j_k (A^{im}_{ml} \! - \! A^{mi}_{lm}) \! - \! \delta^j_l (A^{im}_{mk} \! - \! A^{mi}_{km}) \right] \\
		[\irrep{D}_\textbf{S}]^{ij}_{kl} = & \frac{1}{4} (A^{ij}_{kl} + A^{ji}_{kl} + A^{ij}_{lk} + A^{ji}_{lk}) + \frac{1}{2 (N^2 \! + \! 3N \! + \! 2)} A^{mn}_{nm} \left(\delta^i_k\delta^j_l +  \delta^i_l\delta^j_k\right) \\
		& - \! \frac{1}{4(N \! + \! 2)} \! \left[ \delta^i_k (A^{jm}_{ml} \! + \! A^{mj}_{lm}) \! + \! \delta^i_l (A^{jm}_{mk} \! + \! A^{mj}_{km}) \! + \! \delta^j_k (A^{im}_{ml} \! + \! A^{mi}_{lm}) \! + \! \delta^j_l (A^{im}_{mk} \! + \! A^{mi}_{km}) \right] .
	\end{aligned}
\end{equation}
The above decomposition applies for $N \geq 4$, however, in the case of QCD with $N = 3$ the representation $\irrep{B}_\textbf{S}$ does not appear and the symmetric adjoint representation for $SU(3)$ is given by $[\irrep{8}_\textbf{S}]^{ij}_{kl} = [\irrep{A}_\textbf{S}]^{ij}_{kl} + [\irrep{B}_\textbf{S}]^{ij}_{kl}$. This concludes the decomposition of the amplitudes considered in sections~\ref{sec:decomposing:qcd:cross:section} and~\ref{sec:sun:color:decomposition}.

\subsection{Decuplet annihilation}
\label{sec:exotic:qcd:decuplet}
It is possible to imagine dark sectors with exotic and large representations of $SU(3)$. Out of these representations the $\irrep{10}$, $\irrep{15}$ and $\irrep{27}$ will have annihilations directly into two Standard Model particles, either to two gluons or a quark gluon pair. Although building these models is challenging, large color representations are associated to higher annihilation cross sections compared to the models we study --- see equation~\eqref{eq:analytic:xsec:annihilation}. The large annihilation rate leads to an enhanced depletion of the dark matter relic abundance or equivalently allows for larger mass splittings between the dark matter and its coannihilation partner. As an example we consider the $\irrep{10}$ for which we have the following color decomposition
\begin{equation}
	\irrep{10} \otimes \overline{\irrep{10}} = \irrep{1} \oplus \irrep{8} \oplus \irrep{27} \oplus \irrep{64} .
\end{equation}
By virtue of equation~\eqref{eq:potential:color:decomposition} while inserting the quadratic Casimir invariants from equation~\eqref{eq:casimir:invariants} we decompose the QCD potential as
\begin{equation} \label{eq:potential:decomposition:decuplet}
	V_{\irrep{10} \otimes \overline{\irrep{10}}} = \frac{\alpha_s}{r} \left\{ \begin{aligned} - 6 & \quad (\irrep{1}) \\ - \frac{9}{2} & \quad (\irrep{8}) \\ - 2 & \quad (\irrep{27}) \\ + \frac{3}{2} & \quad (\irrep{64}) \end{aligned} \right. .
\end{equation}

To decompose the total cross section we write the color part of the amplitude in tensor notation as $A^{ijk}_{lmn} = v^{ijk} w_{lmn}$ with full symmetricity in the upper and lower components separately. After doing the calculation we find
\begin{equation} \label{eq:amplitude:decomposition:decuplet}
	\begin{aligned}
		{[\irrep{1}]}^{ijk}_{lmn} & = \frac{1}{60} D^{ijk}_{lmn} \\
		{[\irrep{8}]}^{ijk}_{lmn} & = \frac{1}{30} C^{ijk}_{lmn} - \frac{1}{30} D^{ijk}_{lmn} \\
		{[\irrep{27}]}^{ijk}_{lmn} & = \frac{1}{7} B^{ijk}_{lmn} - \frac{2}{35} C^{ijk}_{lmn} + \frac{3}{140} D^{ijk}_{lmn} \\
		{[\irrep{64}]}^{ijk}_{lmn} & = A^{ijk}_{lmn} - \frac{1}{7} B^{ijk}_{lmn} + \frac{1}{42} C^{ijk}_{lmn} - \frac{1}{210} D^{ijk}_{lmn} .
	\end{aligned}
\end{equation}
In these equations we used
\begin{equation}
	\begin{aligned}
		B^{ijk}_{lmn} & = \delta^i_l A^{pjk}_{pmn} + \cdots \\
		C^{ijk}_{lmn} & = \delta^i_l \delta^j_m A^{pqk}_{pqn} + \cdots \\
		D^{ijk}_{lmn} & = \delta^i_l \delta^j_m \delta^k_n A^{pqr}_{pqr} + \cdots ,
	\end{aligned}
\end{equation}
where the dots represent all symmetric combinations in the upper and lower indices. $B^{ijk}_{lmn}$ has nine terms, $C^{ijk}_{lmn}$ has eighteen terms and $D^{ijk}_{lmn}$ has six terms. By inserting equation~\eqref{eq:amplitude:propto:gg} into equation~\eqref{eq:amplitude:decomposition:decuplet}, we then obtain the following decomposition for the $\irrep{10} \otimes \overline{\irrep{10}} \to g^a \, g^b$ process
\begin{equation}
	\begin{aligned}
		\sum_\mathrm{color} \big| A_{\irrep{10} \otimes \overline{\irrep{10}}} \big|^2 & = 7 \sum_\mathrm{color} \big| [\irrep{1}] \big|^2 + \frac{35}{9} \sum_\mathrm{color} \big| [\irrep{8}] \big|^2 + \frac{5}{3} \sum_\mathrm{color} \big| [\irrep{27}] \big|^2 & \quad \mathrm{even} \,\, l + s & \\
		\sum_\mathrm{color} \big| A_{\irrep{10} \otimes \overline{\irrep{10}}} \big|^2 & = \sum_\mathrm{color} \big| [\irrep{8}] \big|^2 & \quad \mathrm{odd} \,\, l + s & .
	\end{aligned}
\end{equation}
As expected the amplitudes for odd $l+s$ only involve the color-octet channel. Moreover, for even $l+s$ we observe no decomposition into the $\irrep{64}$ as this representation does not appear in the color product of two gluons. From this result and equation~\eqref{eq:potential:decomposition:decuplet} the Sommerfeld corrections are obtained equivalently to equation~\eqref{eq:colored:sommerfeld:corrections} as
\begin{equation}
	\sigma^{(S)}_{\irrep{10} \otimes \overline{\irrep{10}} \to g \, g} = \begin{cases} \frac{1}{7} \sigma^{(S)}_C \left[ 6 \alpha_s \right] + \frac{9}{35} \sigma^{(S)}_C \left[ \frac{9 \alpha_s}{2} \right] + \frac{3}{5} \sigma^{(S)}_C \left[ 2 \alpha_s \right] & \mathrm{even} \,\, l + s \\ \sigma^{(S)}_C \left[ \frac{9 \alpha_s}{2} \right] & \phantom{!} \mathrm{odd} \,\, l + s \end{cases}  .
\end{equation}
We observe here that in contrast to the correction factors for the triplet, sextet and octet in equation~\eqref{eq:colored:sommerfeld:corrections}, the decuplet has positive coupling strengths for all of the Coulomb potentials. This implies an even larger enhancement of the annihilation cross sections and strengthens the effect of the larger Casimir values.

\subsection{Triplet--octet annihilation}
\label{sec:exotic:qcd:triplet:octet}
An interesting possible scenario is a dark sector with two colored particles close in mass to the dark matter particle. For example, one could consider a model with a triplet and an octet of $SU(3)$. In addition to the self-annihilation of each of these particles through the strong interaction, the $\irrep{3}$ and the $\irrep{8}$ could coannihilate to a quark and a gluon. The color of the initial state can be decomposed as
\begin{equation}
	\irrep{3} \otimes \irrep{8} = \irrep{3} \oplus \overline{\irrep{6}} \oplus \irrep{15} .
\end{equation}
We then use equation~\eqref{eq:potential:color:decomposition} and~\eqref{eq:casimir:invariants} to decompose the QCD potential which gives
\begin{equation} \label{eq:potential:decomposition:triplet:octet}
	V_{\irrep{3} \otimes \irrep{8}} = \frac{\alpha_s}{r} \left\{ \begin{aligned} - \frac{3}{2} & \quad (\irrep{3}) \\ - \frac{1}{2} & \quad (\overline{\irrep{6}}) \\ + \frac{1}{2} & \quad (\irrep{15}) \end{aligned} \right. .
\end{equation}

To decompose the total cross section we write the color part of the amplitude in tensor notation as $A^{ij}_k = v^{i} w^j_k$ with the condition $A^{im}_m = 0$. We then find
\begin{equation} \label{eq:amplitude:decomposition:triplet:octet}
	\begin{aligned}
		{[\irrep{3}]}^{ij}_k & = \frac{3}{8} \delta^i_k A^{mj}_m - \frac{1}{8} \delta^j_k A^{mi}_m \\
		{[\overline{\irrep{6}}]}^{ij}_k & = \frac{1}{2} (A^{ij}_k - A^{ji}_k) + \frac{1}{4} (\delta^j_k A^{mi}_m - \delta^i_k A^{mj}_m) \\
		{[\irrep{15}]}^{ij}_k & = \frac{1}{2} (A^{ij}_k + A^{ji}_k) - \frac{1}{8} (\delta^j_k A^{mi}_m + \delta^i_k A^{mj}_m) .
	\end{aligned}
\end{equation}
In this model we assume a coupling between new particles that transform under the $\irrep{3}$ and the $\irrep{8}$ and a Standard Model quark --- such as the squark-gluino coupling in supersymmetry. The $\irrep{3} \otimes \irrep{8} \to q \, g$ coannihilation process can then occur through either an $s$-channel quark, a $t$-channel $\irrep{3}$ or a $t$-channel $\irrep{8}$. These diagrams have different color structures and their relative strength determines the decomposition over the three different color channels. In contrast to the processes considered in the main part an $l+s$ symmetry is not applicable to this process since neither the initial nor the final state involve pairs of identical particles. Therefore a decomposition like
\begin{equation}
	\sum_\mathrm{color} \big| A_{\irrep{3} \otimes \irrep{8}} \big|^2 = \alpha \sum_\mathrm{color} \big| [\irrep{3}] \big|^2 = \beta \sum_\mathrm{color} \big| [\overline{\irrep{6}}] \big|^2 = \gamma \sum_\mathrm{color} \big| [\irrep{15}] \big|^2 .
\end{equation}
will have momentum-dependent factors $\alpha$, $\beta$ and $\gamma$, rendering the calculation of Sommerfeld corrections to be more cumbersome.

A possible approach is based on the fact that each of the three diagrams contributing to the amplitude has a fixed color structure. When squaring the amplitude, the squares of the contributions of each diagram as well the interference terms will also have a well-defined color structure when considered separately. Applying the recipe described in section~\ref{sec:decomposing:qcd:sommerfeld} to each of these terms will give the total analytic Sommerfeld-corrected cross section at a given order in the partial wave expansion. For convenience we present here the color decompositions of each combination of channels:
\begin{equation}
	\begin{aligned}
		& \alpha = 1, \qquad && \beta = 0, \qquad && \gamma = 0 & \qquad s\textrm{-channel squared} \phantom{.} \\
		& \alpha = 64, \qquad && \beta = \frac{32}{9}, \qquad && \gamma = \frac{64}{45} & \qquad t_3\textrm{-channel squared} \phantom{.} \\
		& \alpha = \frac{16}{9}, \qquad && \beta = 8, \qquad && \gamma = \frac{16}{5} & \qquad t_8\textrm{-channel squared} \phantom{.} \\
		& \alpha = 1, \qquad && \beta = 0, \qquad && \gamma = 0 & \qquad s\textrm{-channel interference} \phantom{.} \\
		& \alpha = 8, \qquad && \beta = 4, \qquad && \gamma = \frac{8}{5} & \qquad t_{3/8}\textrm{-channel interference} .
	\end{aligned}
\end{equation}

\subsection{Triplet--triplet annihilation}
\label{sec:exotic:qcd:triplet:triplet}
Here, we consider a model involving a scalar dark matter particle that couples to a Standard Model quark and a new vector-like quark ($\psi$) which is a triplet under color~\cite{Giacchino:2015hvk}. In this case, the annihilation of $\psi$ into quark pairs can occur through two independent processes, namely $\psi \, \bar{\psi} \to q \, \bar{q}$ and $\psi \, \psi \to q \, q$ plus its conjugate. Annihilation into a quark anti-quark pair can occur through either an $s$-channel gluon or a $t$-channel scalar (the dark matter). $\psi$ annihilation to identical quarks occurs either in the $t$ and $u$-channel through dark matter exchange.

First we discuss the effect of the $t$-channel dark matter exchange to $\psi \, \bar{\psi} \to q \, \bar{q}$. The corresponding potential has been derived in~\eqref{eq:decomposed:potentials} and the resulting Sommerfeld correction for the $s$-channel gluon exchange has been derived in section~\ref{sec:decomposing:qcd:cross:section} and presented in equation~\eqref{eq:colored:sommerfeld:corrections}. Including the new $t$-channel diagrams leads to a more complex color structure. As for the triplet-octet model, since neither the initial state nor the final state involve identical particles, the $(l,s)$ components of the amplitude cannot be constrained by symmetry arguments. We have to adopt the same strategy as in section~\ref{sec:exotic:qcd:triplet:octet} and observe that the $t$-channel amplitude decomposes as
\begin{equation}
	\sum_\mathrm{color} \big| A_{\irrep{3} \otimes \bar{\irrep{3}}}^{t\textrm{-channel}} \big|^2 = 9 \sum_\mathrm{color} \big| [\irrep{1}] \big|^2 = \frac{9}{8} \sum_\mathrm{color} \big| [\irrep{8}] \big|^2 ,
\end{equation}
whereas interference between the $s$ and $t$-channel only occurs when the initial state is in the octet representation.

The situation changes for the second process $\psi \, \psi \to q \, q$. Since the quarks in the final state are identical, contributions from each $(l,s)$ state are constrained by symmetry. The color part of this amplitude can be decomposed as
\begin{equation}
	\irrep{3} \otimes \irrep{3} = \overline{\irrep{3}} \oplus \irrep{6} .
\end{equation}
We then use equations~\eqref{eq:potential:color:decomposition} and~\eqref{eq:casimir:invariants} to decompose the QCD potential which gives
\begin{equation} \label{eq:potential:decomposition:triplet:triplet}
	V_{\irrep{3} \otimes \irrep{3}} = \frac{\alpha_s}{r} \left\{ \begin{aligned} - \frac{2}{3} & \quad (\irrep{3}) \\ + \frac{1}{3} & \quad (\overline{\irrep{6}}) \end{aligned} \right. .
\end{equation}
To decompose the total cross section we write the color part of the initial states in tensor notation as $A^{ij} = v^i w^j$ and find
\begin{equation} \label{eq:amplitude:decomposition:triplet:triplet}
	\begin{aligned}
		{[\overline{\irrep{3}}]}^{ij} & = \frac{1}{2} (A^{ij} - A^{ji}) \\
		{[\irrep{6}]}^{ij} & = \frac{1}{2} (A^{ij} + A^{ji}) .
	\end{aligned}
\end{equation}
The total color structure for this process can be written as the sum of the $t$-channel and $u$-channel contributions, namely $A^{ij}_{kl} = \alpha \delta^i_k \delta^j_l + \beta \delta^j_k \delta^i_l$. The CP symmetry condition akin to that of equation~\eqref{eq:color:amplitude:symmetricity} imposes that $\alpha = \beta$ for $l + s$ even and $\alpha = - \beta$ for $l + s$ odd. Inserting this information into equation~\eqref{eq:amplitude:decomposition:triplet:triplet} gives the following color decomposition
\begin{equation}
	\begin{aligned}
		\sum_\mathrm{color} \big| A_{\irrep{3} \otimes \irrep{3}} \big|^2 & = \sum_\mathrm{color} \big| [\irrep{6}] \big|^2 & \quad \mathrm{even} \,\, l + s & \\
		\sum_\mathrm{color} \big| A_{\irrep{3} \otimes \irrep{3}} \big|^2 & = \sum_\mathrm{color} \big| [\overline{\irrep{3}}] \big|^2 & \quad \mathrm{odd} \,\, l + s & ,
	\end{aligned}
\end{equation}
which, using equation~\eqref{eq:potential:decomposition:triplet:triplet}, leads to
\begin{equation}
	\sigma^{(S)}_{\irrep{3} \otimes \irrep{3} \to q \, q} = \begin{cases} \sigma^{(S)}_C \left[ - \frac{\alpha_s}{3} \right] & \mathrm{even} \,\, l + s \\ \sigma^{(S)}_C \left[ \frac{2 \alpha_s}{3} \right] & \phantom{!} \mathrm{odd} \,\, l + s \end{cases}
\end{equation}
which, beyond the $s$-wave, contrasts with the result derived in~\cite{Giacchino:2015hvk} for a similar simplified model.

\bibliographystyle{JHEP}
\bibliography{sommerfeld_v4}

\end{document}